\documentclass[10pt,aps,pop,showpacs,superscriptaddress,floatfix,reprint,twocolumn
]{revtex4-1} 
\usepackage[squaren, Gray, cdot]{SIunits}
\usepackage{array}
\newcolumntype{H}{>{\setbox0=\hbox\bgroup}c<{\egroup}@{}}
\usepackage {revsymb}
\usepackage{bm}

\usepackage{amssymb,amsmath}
\usepackage{graphicx}
\usepackage{float}
\usepackage{bm}
\usepackage{rotating}
\usepackage{multirow}
\usepackage{dcolumn}
\usepackage{tabularx}
\usepackage{tabulary}
\usepackage{longtable}
\usepackage{hyperref}
\usepackage{cleveref}
\usepackage[pdftex,dvipsnames,usenames]{color}

\usepackage{color, colortbl}
\definecolor{LightCyan}{rgb}{0.88,1,1}
\definecolor{RedLink}{rgb}{0.47,0,0}

\usepackage{rotating}

\newlength{\dbarheight}

\begin{document}

\title{Polarity-field driven conductivity in SrTiO$_3$/LaAlO$_3$: a hybrid functional study}

\author{S\'ebastien Lemal}
\affiliation{Physique Th\'eorique des Mat\'eriaux, Q-MAT, CESAM, Li\`ege
  (B5), B-4000 Li\`ege, Belgium}
\author{Nicholas C. Bristowe}
\affiliation{Department of Materials, Imperial College London, Exhibition Road, London, SW7 2AZ, United Kingdom}
\affiliation{School of Physical Sciences, University of Kent, Canterbury CT2 7NH, United Kingdom}
\author{Philippe Ghosez}
\affiliation{Physique Th\'eorique des Mat\'eriaux, Q-MAT, CESAM, Li\`ege
  (B5), B-4000 Li\`ege, Belgium}
\date{\today}
\setcounter{page}{1} 

\newcommand{\ls}[1]{{\textcolor{red}{#1}}}
\newcommand{\phg}[1]{{\textcolor{blue}{#1}}}
\newcommand{\ncb}[1]{{\textcolor{magenta}{#1}}}
\newcommand{\mjv}[1]{{\textcolor{green}{#1}}}
\newcommand{\dib}[1]{{\textcolor{cyan}{#1}}}

\begin{abstract}
The origin of the 2-dimensional electron system (2DES) appearing at the (001) interface of band insulators $\rm SrTiO_3$ and $\rm LaAlO_3$ has been rationalized in the framework of a polar catastrophe scenario. This implies the existence of a critical thickness of polar $\rm LaAlO_3$ overlayer ($4~\rm u.c.$) for the appearance of the 2DES: polar catastrophe for thick $\rm LaAlO_3$ overlayer is avoided either through a Zener breakdown or a stabilization of donor defects at the $\rm LaAlO_3$ surface, both providing electrons to dope the substrate. 
The observation of a critical thickness is observed in experiments, supporting these hypotheses. Yet, there remains an open debate about which of these possible mechanisms actually occurs first. Using hybrid functional Density Functional Theory, we re-examine these mechanisms at the same level of approximation. Particularly, we clarify the role of donor defects in these heterostructures, and argue that, under usual growth conditions, electric-field driven stabilization of oxygen vacancies and hydrogen adsorbates at the LAO surface occur at a smaller LAO thickness than required for Zener breakdown.
\end{abstract}


\maketitle


\section{Introduction}

Functionalities offered by oxides compounds (TMOs), related to their electronic structures, sparked tremendous interest for technological applications. Moreover, further interests has been attracted by interfaces between TMO compounds, which exhibit emerging properties not present in either parent compounds at the bulk level. One of the most studied emerging property is 2-dimensional conductivity at interfaces between wide-gap insulator TMOs, which has been originally observed at the (001) interface between $\rm LaAlO_3$ (LAO) and $\rm SrTiO_3$ (STO). The origin of the observed conductivity is mainly attributed to a 2-dimensional electron system (2DES). In addition to conductivity, several other properties have been attributed to this 2DES, such as superconductivity, magnetism, confinement effects, \emph{etc}~\cite{gariglio2009, zubko2011,hwang2012,banerjee2013,salluzzo2013,chakhalian2014, khomskii2014, gariglio2016}. 
The appearance of the 2DES has been attributed to a polar discontinuity~\cite{ohtomo2004} between between STO and LAO. Hence, growing a LAO overlayer on top of a STO(001) substrate can trigger conductivity as long as the LAO film is thicker than 3 units cells (u.c.). This critical thickness is though to originate from to the presence of an electric field in the LAO layer: for a sufficiently thick overlayer, charge transfer occurs to avoid a divergence of the electrostatic potential.

In spite of considerable research, there remains a debate around the origin of the 2DES. The polarity-driven mechanism has been formulated either as a purely electronic reconstruction~\cite{ohtomo2004,savoia2009} (or Zener breakdown), or as a polarity-driven stabilization of oxygen vacancies (or other donor defects) at the LAO surface~\cite{li2011, bristowe2011,yu2014}, providing electrons which remain confined near the interface. Nevertheless, the existence of an electric field in the LAO layer remains debated, although some experiments support this~\cite{cancellieri2011, reinleschmitt2012, li2018}. Other hypotheses include as-grown oxygen deficiency, off-stoichiometry, or surface adsorption~\cite{kim2014, scheiderer2015, brown2016, zhang2018}. It remains unclear how these mechanisms coexist or dominate.

In this study we re-examine, using a combination of first-principles calculations and phenomenological models, the electric-field driven mechanisms at the origin of the 2DES. The different models are analyzed in turn and explored through hybrid functional DFT calculations. We then compare them and rationalize experimental findings obtained from STO/LAO heterostructures. A specific focus is given to the surface redox model, and we discuss how the tunability of the crystal thickness obtained by alloying the polar overlayer is rationalized within this model. 


\section{Technical details}

\begin{figure}
	\centering
		\includegraphics[width=.495\textwidth]{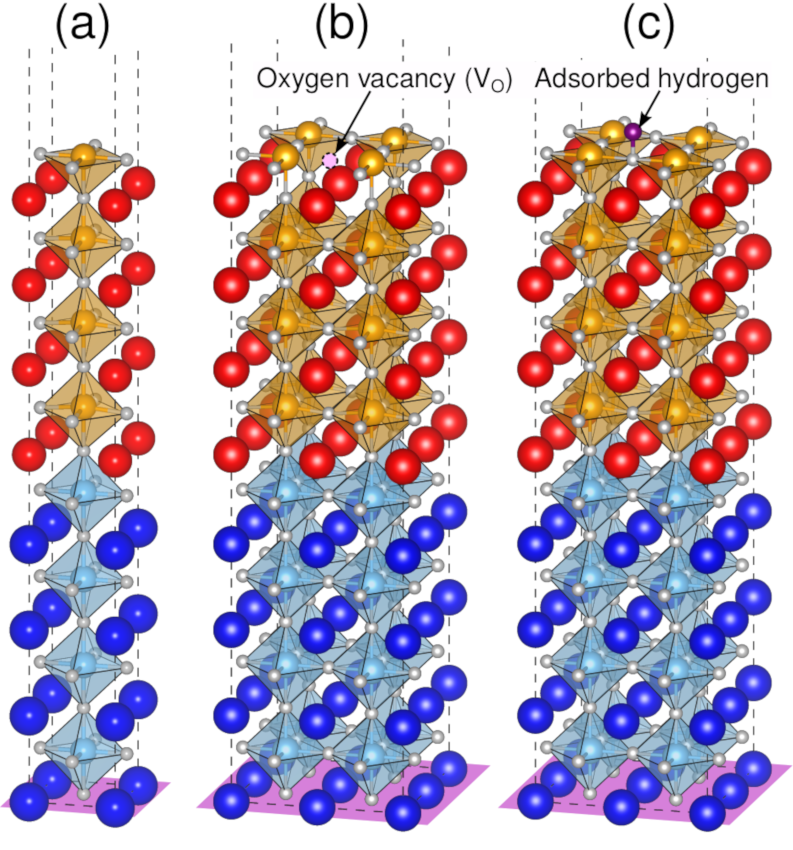}
        \caption{Examples of simulation cells used for calculations: (a) a $1\times 1$ simple cell, (b) a $2\times 2$ supercell with one $\rm V_O$ at the LAO surface and (c) a $2\times 2$ supercell with one H atom adsorbed on the LAO surface. In the last two examples, the area density of $\rm V_O$ or adsorbed H is $\eta = 1/4a^2 = 1/4\square$. $2 \times 3$ supercells have also been used in this study (not shown). The films are symmetric, the symmetry plane is shown in purple.}
        \label{fig:HS}
\end{figure}

Before going further, we specify the technical details of our first-principles calculations. We use the CRYSTAL code~\citep{refcrystal14} to compute from DFT the atomic and electronic structure of bulk $\rm LaAlO_3$ and $\rm SrTiO_3$ systems, as well as heterostructures based on these compounds. Examples of simulation cell in slab geometry used in this study are shown in Fig.~\ref{fig:HS}. 

A Gaussian basis set was used to represent the electrons. All the electrons have been included for Ti~\cite{bredow2004}, O~\cite{piskunov2004} and Al~\cite{SBSNbasis}, while we use a Hartree-Fock pseudopotential~\cite{piskunov2004} for Sr and the Stuttgart energy-consistent pseudopotential~\cite{cao2004} for La. The basis sets of Sr and O have been optimized for STO. In the basis set of La, the Gaussian exponents smaller than $0.1$ were disregarded and the remaining outermost polarization exponents for the $10s$, $11s$ shells (0.5672, 0.2488), $9p$, $10p$ shells (0.5279, 0.1967), and $5d$, $6d$, $7d$ shells (2.0107, 0.9641, 0.3223), together with Al $4sp$ (0.1752) exponent from the 8-31G Al basis set, were optimized for LAO.

The exchange-correlation energy is modelled with the B1-WC hybrid functional~\cite{dbilcb1wc}{, which have been used in several previous studies about STO, LAO and their interfaces~\cite{cancellieri2011,delugas2011,reinleschmitt2012,cancellieri2014,bilc2016,li2018,caputo2020}}.
A Monkhorst-Pack mesh~\cite{monkhorstpack1972} of $6 \times 6 \times 6$ special $k$-points is used for cubic bulk LAO and STO, ensuring a proper convergence of the total energy below 1 meV per formula unit. The sampling is then refined into a $12 \times 12 \times 12 $ mesh of special $k$-points for the computation of properties such as the electronic density of states (DOS) or the vibrational modes at the $\Gamma$ point in the irreducible Brillouin Zone (IBZ).

{Concerning the heterostructures, different sizes of simulation cells have been used, shown in Fig.~\ref{fig:HS}.(b) and .(c), have been used in the study ($2 \times 2$ and $2 \times 3$ supercells), with adapted Brillouin zone sampling with respect to supercell size. For the $1 \times 1$ simulation cell (see Fig~\ref{fig:HS}a) and the  $2 \times 2$ simulation cell (see Fig \ref{fig:HS}b) the Brillouin Zone sampling is adapted to a $6 \times 6 \times 1$ mesh and $3 \times 3 \times 1$ respectively. It is then refined to $12 \times 12 \times 2$ (respectively $6 \times 6 \times 2$) to compute the electronic band structure and related DOS.} A  smearing of the Fermi surface has been set to $k_B T = 0.001~\rm Ha$. The self-consistent DFT cycles are considered to be converged when the energy change between cycles are smaller than $\rm 10^{-8}~Ha$. The optimization of the atomic positions are performed with convergence criteria of $1.5 \times 10^{-4}\rm~Ha/Bohr$ in the root-mean square values of the energy gradients, and $1.2 \times 10^{-3}\rm~Bohr$ in the root-mean square values of the atomic displacements. The evaluation of the Coulomb and exchange series is determined by five parameters, fixed to their default~\cite{refcrystal14manual} values: 7, 7, 7, 7 and 14.

We consider the effect of oxygen vacancies, acting as double donors according to the following surface redox reaction:
\begin{eqnarray}
\rm O^{2-} \rightarrow \frac{1}{2}\:O_2 + V_O + 2\:e^-
\end{eqnarray}
The technical details for the calculations of the systems with oxygen vacancies are similar to the ones used for the pristine slabs in term of basis sets, convergence threshold and investigated geometrical configurations for the heterostructures.
Oxygen vacancies ($\rm V_O$) have been modelled by removing explicitly an oxygen atom from its site (core and electrons), while leaving ``ghost" oxygen basis functions on the site to properly model the electron density within the vacancy.

Formation energies $E_{f,\mu = 0}$ are calculated in the $\rm O$-rich limit from the relation:
\begin{eqnarray}
E_{f,\mu = 0} &=& \frac{1}{n_v}\left[ E_V - [E_0 - n_v \frac{1}{2} E_{\rm O_2}]\right]\label{eqn:formation_energy_formula}
\end{eqnarray}
where $E_V$ and $E_0$ are the calculated total energy of the systems with and without $\rm V_O$ (and same cell size), $n_v$ the number of $\rm V_O$ in the supercell, and $E_{\rm O_2}$ the calculated total energy of the single $\rm O_2$ molecule in the triplet state. 

Equation~\ref{eqn:formation_energy_formula} only considers the enthalpic contribution at 0~K to the formation energies. To account for the atmosphere during growth at finite temperature and pressure, considering the environment as a reservoir, one has to consider the chemical potential of oxygen $\mu_{\rm O}(p,T)$ relative to the gaseous phase at finite oxygen partial pressure $p$ and temperature $T$, such:
\begin{eqnarray}
\displaystyle E_{f,\mu} &=& \frac{1}{n_v}\left[ E_V - [E_0 - n_v \frac{1}{2} (E_{\rm O_2} + \mu_{\rm O_2)}(p,T)]\right]\label{eqn:formation_energy_formula_mu1}\\
\displaystyle E_{f,\mu} &=& E_{f,\mu = 0} + \mu_{\rm O}(T, p_{\rm O_2})\label{eqn:formation_energy_formula_mu2}
\end{eqnarray}
In the relation above, $\mu_{\rm O}(p,T)$ is calculated from the thermodynamic model~\cite{reuter2001,osorio-guillen2006}: the details of the calculation can be found in Appendix~\ref{AppendixA}. $\mu_{\rm O}(p,T)$ is usually considered as a parameter depending on the environment; for the purpose of this study, we consider $\mu_O = -2~\rm eV$ according to the growth conditions of standard STO/LAO heterostructures (identical to the value used in Reference~\onlinecite{yu2014}) and consider variations due to growth according to the thermodynamical model (Appendix~\ref{AppendixA}).

Finally, we also consider the surface redox hydroxylation process, where H atoms are adsorbed at the LAO surface. The chemical reaction of this process is written as:
\begin{eqnarray}
{\rm 2\:O^{2-} + H_2O \rightarrow \frac{1}{2}O_2 + 2\:OH^- + 2\:e^-}\label{eqn:hydroxylation_reaction}
\end{eqnarray}
This process provides 1 donor electron per hydroxyl $\rm OH^-$ (in contrast to the non-redox hydroxylation process, $\rm H_2O \rightarrow H^+ + OH^-$, which does not provide any carriers). Hence, the adsorption energy of hydrogen according to the process of Eqn.~\eqref{eqn:hydroxylation_reaction} is calculated from DFT, in the supercell approach, as:
\begin{eqnarray}
\displaystyle E_{A,\mu} &=& \frac{1}{n_A}\left( E_A + \frac{n_A}{4} [E_{\rm O_2} + \mu_{\rm O_2}(T, p_{\rm O_2})]\right)\nonumber \\
& &  - \frac{1}{n_A}\left( E_0 + \frac{n_A}{2}[E_{\rm H_2O} + \mu_{\rm H_2O}(T, p_{\rm H_2O})]\right)\label{eqn:formation_energy_H_adsorbate}
\end{eqnarray}
where $n_A$ is the number of adsorbed H atoms in the cell, $E_A$ the total energy of the heterostructure with the adsorbed H atoms, $E_{\rm H_2O}$ the total energy of a water molecule, and $\mu_{\rm H_2O}$ the chemical potential of water in gaseous form, calculated from the thermodynamic model. All the other quantities in Eqn.~\ref{eqn:formation_energy_H_adsorbate} are the same as defined earlier.


\section{Results} 

We will now provide a description of the electric-field driven hypotheses at the origin of the 2DES at the (001) STO/LAO interface, the Zener breakdown and the surface oxygen vacancies. The mechanisms can be explained in the framework of a polar catastrophe, where the diverging electrostatic potential in the LAO film is the driving force behind the instability leading to the appearance of the 2DES. The main argument in favor of these mechanisms is the existence of threshold LAO thicknesses to witness different phenomena, such as signatures of $\rm Ti^{+3.5}$ valence at the interface from spectroscopy, or the change in sheet resistance. Indeed, as will be argued in the following discussion, the intricacies of the different mechanisms result in differences in properties.

The parameters of each model are evaluated through hybrid functional DFT, using the B1-WC hybrid functional. This functional predicts good properties for bulk STO and LAO, and has been used extensively to study the STO/LAO interface. As an example, we mention the electronic band gap of cubic STO and rhombohedral LAO as calculated from B1-WC: $E_{g}^{\rm STO,c} = 3.56~{\rm eV}$ and $E_{g}^{\rm LAO,r} = 5.78~{\rm eV}$, showing good agreement with experiments ($3.25~\rm eV$~\cite{vanbenthem2001} and $5.6~\rm eV$~\cite{lim2002} respectively). In addition, the B1-WC hybrid functional predicts with good accuracy the dielectric constant of LAO, which is an important parameter in all explored models: $\varepsilon^{\rm LAO,c}_{r} = 27$ for cubic LAO and $\varepsilon^{\rm LAO,c}_{r} = 21$ for rhombohedral LAO. 

\subsection{Electric-field driven Zener breakdown}

For the STO/LAO interface, it is \emph{a priori} possible for the electrons to rearrange themselves to avoid a polar catastrophe, as the electrostatic potential diverges with increasing LAO thickness. This is the so-called Zener breakdown scenario, and does not involve any atomic reconstruction, since only the electronic population changes. In this Section, we will focus on the Zener breakdown hypothesis and its description from first-principles calculations based on the hybrid functional formalism, with the B1-WC hybrid functional. This will benchmark our results based from the different hypotheses.

The Zener breakdown stems from the electrostatic behavior of the STO/LAO interface and can be formulated in terms of the conservation of the normal component of the displacement field $\mathbf{D}$ across the interface~\cite{stengel2009}. In the (001) direction, the LAO layers can be considered as a serie of capacitors with $\rm (LaO)^{1+}$ and $\rm (AlO_2)^{1-}$, corresponding to surface charge $\sigma_0^{\rm LAO} = 0.5~e/\square$, where $e$ is the electron charge and $\square$ is the in-plane unit cell area ($\square = a^2$). Hence, considering the polarity of each LAO monolayer, the LAO film has a formal polarization of $P_0^{\rm LAO} = -e/2\square$. 
As the STO atomic planes are neutral, there is no formal polarization in the STO substrate. The formal polarizations of STO and LAO are therefore:
\begin{eqnarray}
P_{0}^{\rm STO} &=& 0\\\label{eqn:formal_pol_STO}
P_{0}^{\rm LAO} &=& -e/2\square\label{eqn:formal_pol_LAO}
\end{eqnarray}
The transverse component of the displacement field, in each environment (STO, LAO, vacuum) is then:
\begin{eqnarray}
D^{\rm STO} &=& \varepsilon_{0}E_{0}^{\rm STO} + P_{0}^{\rm STO}\\\label{eqn:displacement_STO}
D^{\rm LAO} &=& \varepsilon_{0}E_{0}^{\rm LAO} + P_{0}^{\rm LAO}\\\label{eqn:displacement_LAO}
D^{\rm vac} &=& 0\label{eqn:displacement_vacuum}
\end{eqnarray}
In the absence of free charges, which is the case for band insulators, the normal component of the displacement field has to be preserved~\cite{vanderbilt1993}. Hence, the vacuum fixes $D = 0$ across the whole heterostructure, and an electric field appears in the LAO overlayer, such that:
\begin{eqnarray}
E_{0}^{\rm STO} &=& -\frac{P_{0}^{\rm STO}}{\varepsilon_0} = 0 \\\label{eqn:el_field_STO}
E_{0}^{\rm LAO} &=& -\frac{P_{0}^{\rm LAO}}{\varepsilon_0} = \frac{1}{\varepsilon_0}\frac{e}{2\square}\label{eqn:el_field_LAO}
\end{eqnarray}
Since $\rm LaAlO_3$ is an insulator, the material will polarize under the effect of an electric field, leading to a depolarizing field and surface induced bound charges $\sigma^{\rm LAO}_{ind}$. 
The polarization induced in LAO is therefore screened by the depolarizing field $E^{\rm LAO}_{0}$ by inducing a dielectric contribution opposite the the formal polarization. This screening depends on the dielectric constant ($\varepsilon_{r}^{\rm LAO} \sim 24$).

\begin{figure*}
	\centering
		\includegraphics[width=.9\textwidth]{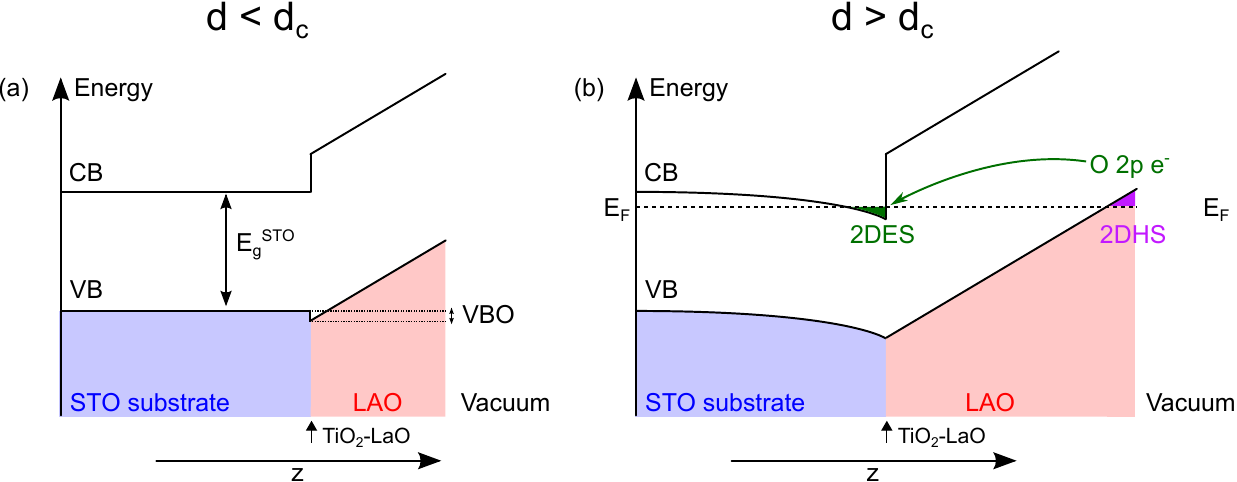}
        \caption[Band diagram representation of the Zener breakdown scenario]{Band diagram representation of the Zener breakdown scenario (a) for a LAO thickness $d$ below the critical thickness $d_c$, (b) for a LAO thickness $d$ above the critical thickness $d_c$.}
        \label{fig:LAOSTO_ntype_band_diagram}
\end{figure*}

The resulting electric field $E^{\rm LAO}$ and surface charge are given by:
\begin{eqnarray}
E^{\rm LAO} &=& \displaystyle E^{\rm LAO}_{0} - E^{\rm LAO}_{\rm ind}\nonumber\\
 &=& \frac{1}{\varepsilon_0\varepsilon_{r}^{\rm LAO}}\frac{e}{2\square} = 0.25~\rm V/\angstrom\\\label{eqn:screened_el_field_LAO}
\sigma^{\rm LAO}_{\rm ind} &=& E^{\rm LAO}\varepsilon_0 = 0.02~e/\square\\\label{eqn:LAO_charge}
\sigma^{\rm LAO} &=& \sigma_{0}^{\rm LAO} - \sigma^{\rm LAO}_{\rm ind} = 0.48~e/\square\label{eqn:LAO_induced_charge}
\end{eqnarray}
Within this model, the built-in electric field is estimated to be equal to $0.25~\rm V/\angstrom$. Consequently, the electrostatic potential increases linearly with LAO thickness, by about $c\times E^{\rm LAO} \approx 0.9~\rm eV$ per monolayer ($c$ being the out-of-plane lattice parameter of LAO). This effect can also be viewed in a band diagram, where the valence states of LAO are raised to higher energy with the electrostatic potential, as shown in Fig.~\ref{fig:LAOSTO_ntype_band_diagram}.(a). For a LAO thickness $d$ above a threshold value $d_c$, the valence ${\rm O}~2p$ states at the surface of LAO are raised above the STO conduction band minimum, and a charge transfer occurs from the ${\rm O}~2p$ to the ${\rm Ti}~t_{2g}$ states of STO: a 2DES appears at the interface, as shown in Fig.~\ref{fig:LAOSTO_ntype_band_diagram}.(b); as a by-product of the charge transfer, a 2-dimensional hole system (2DHS) is expected to exist at the surface of LAO according to the Zener breakdown picture. Further insights on the electronic reconstruction and reformulation of the model can be found in Ref.~\onlinecite{bristowe2014}. 

\begin{figure*}
	\centering
		\includegraphics[width=1.\textwidth]{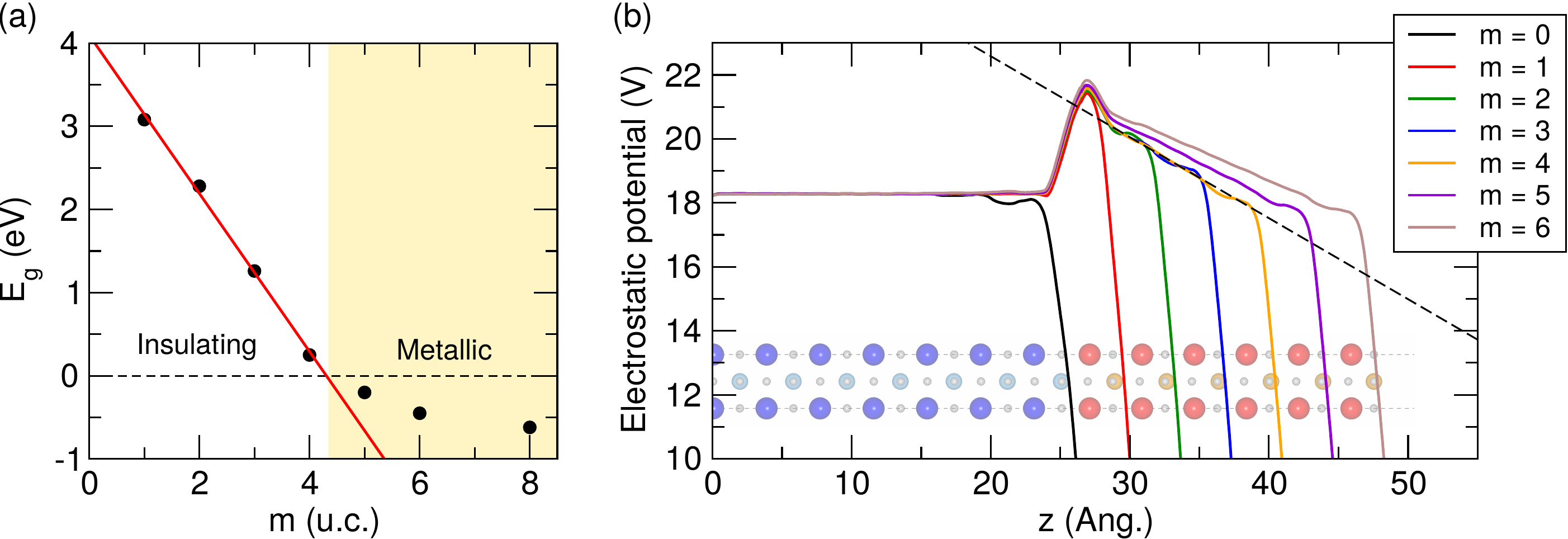}
        \caption[(a) Electronic band gap for ${\rm STO/LAO}_m$/vacuum heterostructures, for different LAO thicknesses and (b) associated macroscopic average of electrostatic potentials]{(a) Electronic band gap for ${\rm STO/LAO}_m$/vacuum heterostructures, for different LAO thicknesses $m$ (u.c.), calculated as the difference between the bottom ${\rm Ti}~t_{2g}$ band and the top of the LAO ${\rm O}~2p$ band. The negative values corresponds to metallic phases; (b) Macroscopic average of the electrostatic potential in a ${\rm STO/LAO}_m$ for varying LAO overlayer thicknesses $m$. The slope of electrostatic potential in the LAO layer is estimated to be $-0.25~{\rm V}/\angstrom$ below the onset for Zener breakdown. Above this threshold ($m > 4~\rm u.c.$), the slope decreases with increasing LAO thickness, as the interface progressively gets doped.}
        \label{fig:POTC_slab7_m}
\end{figure*}

\begin{sidewaysfigure}
	\centering
		\includegraphics[width=0.95\columnwidth]{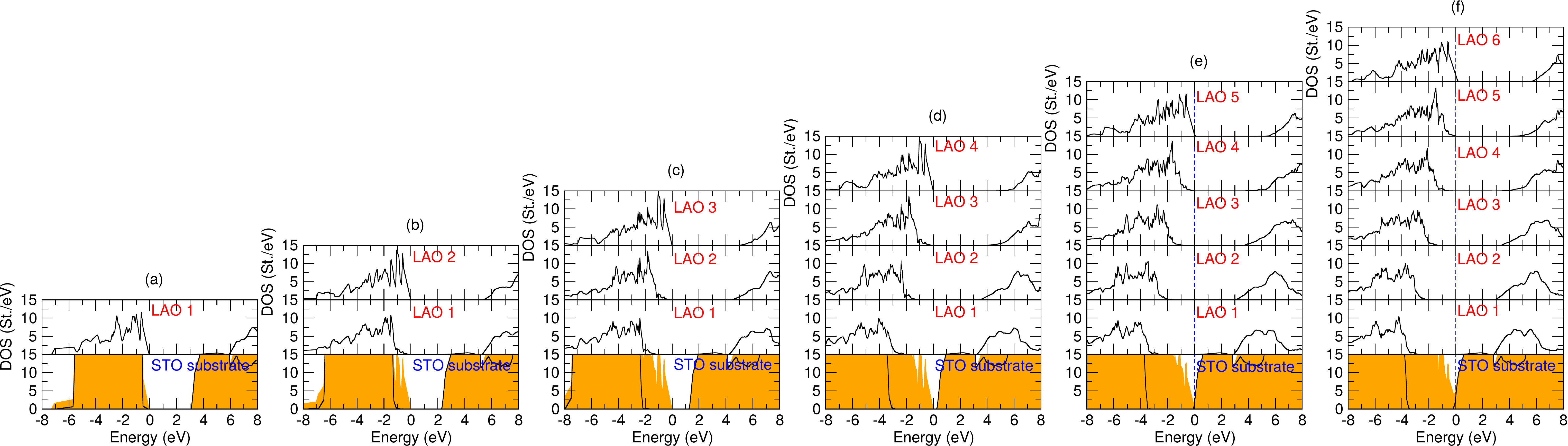}
        \caption[Layer-resolved DOS of ${\rm STO/LAO}_m$ heterostructures]{Layer-resolved DOS of ${\rm STO/LAO}_m$ heterostructures, for varying LAO overlayer thicknesses $m$. The orange area is the total DOS. The vertical dashed line is the Fermi level for the metallic films.}
        \label{fig:LAOSTO_slab7_m_DOSS_layer_resolved}
\end{sidewaysfigure}

First-principles calculations have been performed on ${\rm STO/LAO}_m$ pristine heterostructures. Fig.~\ref{fig:POTC_slab7_m}.(a) shows the evolution of the heterostructure electronic band gap with respect to the number of LAO monolayers {$m$}, and in Fig.~\ref{fig:POTC_slab7_m}.(b) the profile of the macroscopic average of the electrostatic potential across the heterostructures. From these results, we make the following observations: i) the electrostatic potential in the STO is flat, as expected from our earlier considerations; ii) the electrostatic potential varies linearly with increasing LAO thickness, the slope is estimated to be $-0.25~\rm V/\angstrom$ for $1 \leq m \leq 4$, and this results in the linear decrease of the band gap with increasing LAO thickness, with a slope of $-0.9~\rm eV/u.c.$. This is in agreement with the value calculated for a LAO cell with a tetragonal constraint ($a = b = a^{\rm STO}$): $\varepsilon_{r}^{\rm LAO} \approx 24$, in between the value calculated for cubic and rhombohedral LAO (respectively 27 and 21); iii) the field in LAO is expected to raise the valence states in the LAO system, which is shown in the layer-resolved DOS for the different heterostructures (Fig.~\ref{fig:LAOSTO_slab7_m_DOSS_layer_resolved}); and iv) for $m \geq 5~\rm u.c.$, the system is metallic, and for the metallic phases, the slope of the electrostatic potential decreases with increasing LAO thickness. A metal-insulator transition (MIT) is expected to occur at $m = 4.2~\rm u.c.$ based on the linear projection of the evolution of the band gap below the onset for charge transfer. This is the critical thickness of LAO at which an electronic reconstruction occurs (also referred to as a Zener breakdown in the literature). It is also the onset above which the LAO ${\rm O}~2p$ valence states overlap with the ${\rm Ti}~t_{2g}$ conduction states of the STO substrate in the DOS:
\begin{eqnarray}
d_{c}^{\rm ZB} &=& 4.2~\rm u.c.\label{eqn:zener_breakdown_dc}
\end{eqnarray}
The Zener breakdown occurs when the drop of electrostatic potential $\Delta$ across the LAO film is equal to the sum of the band gap of STO, $E_{g}^{\rm STO}$, and the valence band offset VBO, as shown in Fig.~\ref{fig:LAOSTO_ntype_band_diagram}.(a). Hence, it is possible to calculate $d_{c}^{\rm ZB}$ from the Zener breakdown model, by estimating the thickness needed to reach a potential drop equal to $\Delta$ if the slope of the potential is a constant field $E^{\rm LAO}$:
\begin{eqnarray}
d_{c}^{\rm ZB} &=& \frac{\Delta}{E^{\rm LAO}} = \varepsilon_{0}\varepsilon_{r}^{\rm LAO}\frac{\Delta}{P_{0}^{\rm LAO}}\label{eqn:zener_breakdown_dc_el}
\end{eqnarray}
which predicts the same value as in Equation~\ref{eqn:zener_breakdown_dc} by taking the following values, {calculated from DFT on the bulk compounds,} $c = 3.79~\angstrom$, $E_{g}^{\rm STO} = 3.57~\rm eV$ and $\varepsilon_{r}^{\rm LAO} = 24$. The critical thickness $d_{c}^{\rm ZB}$ depends on different physical parameters: the electronic band gap of STO, the valence band offset, the dielectric constant of LAO and the LAO formal charges, which are all intrinsic parameters to the system. Our first-principles calculations predicts $d_{c}^{\rm ZB}$ between 4 and 5 monolayers of LAO. This is an overestimation if we compare to the experiments, for which the onset for conductivity is between 3 and 4~u.c.~\cite{berner2010, berner2013}. 

This overestimation may be attributed to the overestimation of the band gap: all things being equal, correcting the value of the band gap by the experimental one (3.25~eV~\cite{vanbenthem2001}), the critical thickness becomes $d_{c}^{\rm ZB} = 3.8~\rm u.c.$, in better agreement with the experiments.
{We must stress that, as the parameters of the model are sensitive to the methodology (specifically the approximation for the exchange-correlation energy), then the critical thickness as determined from DFT is also sensitive: LDA and GGA severely underestimate the STO band gap, hence the predicted critical thickness reported from DFT studies based on these functionals is slightly underestimated compared to the 4~u.c. value (for example, 3~u.c. in Ref.~\onlinecite{son2009}). Correction to the band gap error using an on-site Hubbard-like $U$ correction~\cite{lee2008-2} eventually fixes this. Another solution to the shortcomings of LDA/GGA consists of using hybrid functional for the exchange-correlation term, such as the one used in this study (B1-WC). Another example, HSE~\cite{yu2014} yields band gap of $\sim$3~eV and a similar critical thickness as the one determined in this study, 4.3~u.c.}

\begin{figure}
	\centering
		\includegraphics[width=1.\columnwidth]{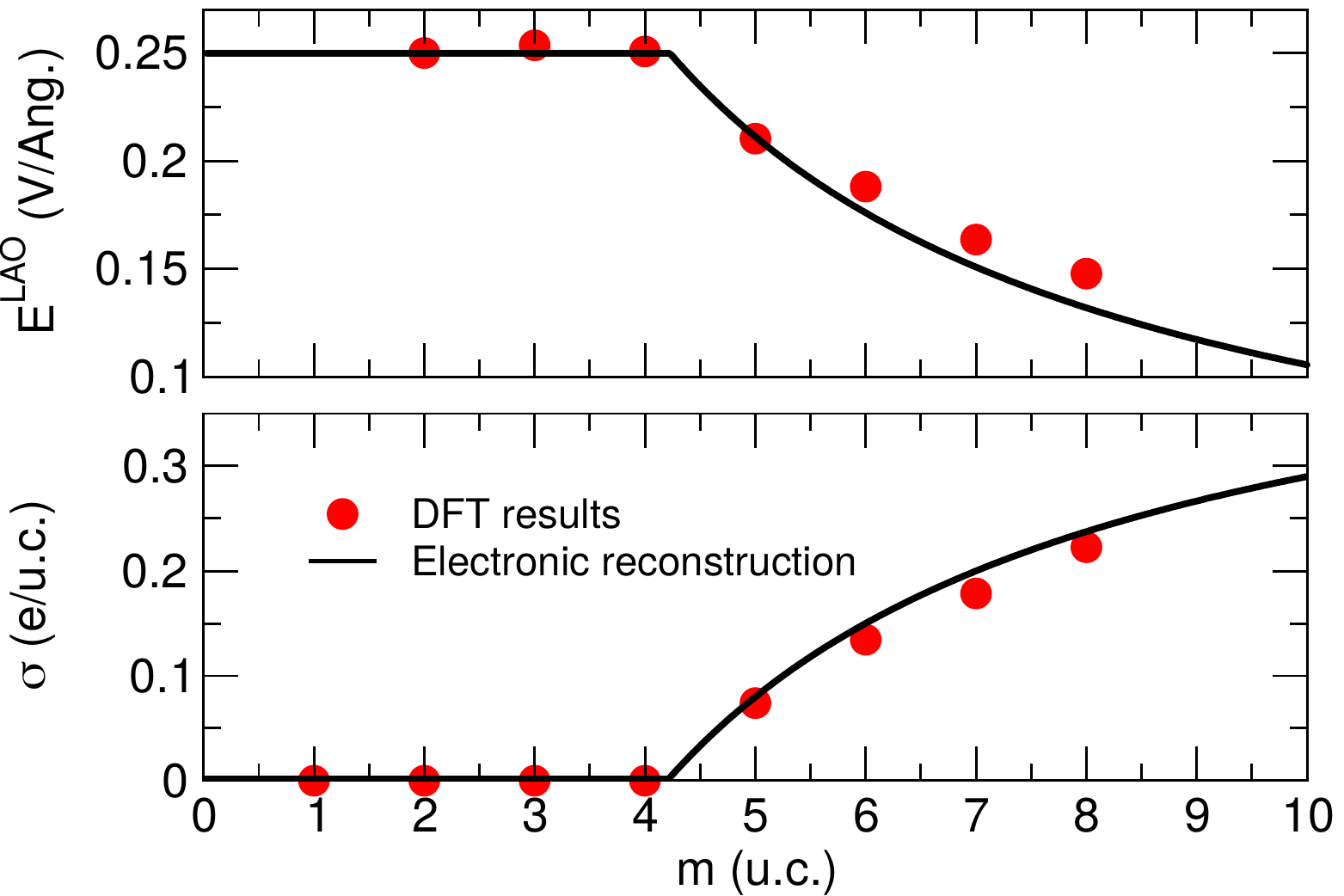}
        \caption{Comparison between the Zener breakdown model (the parameters set in the main text) and the DFT results, for the electric field in the LAO overlayer and the charge density within the STO substrate.}
        \label{fig:ER_vs_DFT}
\end{figure}

Above the critical thickness, each additional LAO layer have their valence electrons at higher energies than the bottom of the conduction band of STO. These electrons are therefore transferred from the surface to the interface.
This transfer pins the valence band edge of the LAO system to the conduction band minimum of STO, which implies that the field in LAO is modified by the charge transfer, according to the following law:
\begin{eqnarray}
E^{\rm LAO} = \frac{\Delta}{d^{\rm LAO}}\label{eqn:field_evolution_with_LAO_thickness}
\end{eqnarray}
Assuming that $\Delta$ is constant, the field in LAO decreases as $1/d^{\rm LAO}$ as the LAO thickness increases. Additionally, the electron transfer leads to the appearance of a 2DES system in the STO subsystem as shown in Fig.~\ref{fig:LAOSTO_ntype_band_diagram}, with a sheet charge density $\sigma_s$~\cite{son2009} as calculated from Equations~\eqref{eqn:field_evolution_with_LAO_thickness}, \eqref{eqn:zener_breakdown_dc_el} and~\eqref{eqn:screened_el_field_LAO}.
\begin{eqnarray}
\sigma_s &=& \frac{1}{2}\frac{e}{\square}\left(1 - \frac{d_{c}^{\rm LAO}}{d^{\rm LAO}}\right)
\end{eqnarray}
The sheet carrier density depends on the dielectric constant of LAO, the thickness of the LAO overlayer and the band gap of STO. In the limit of an infinitely thick LAO overlayer, $\sigma_s$ converges to $\sigma_{0}^{\rm LAO} = 0.5~e/\square$. The evolution of the built-in field $E^{\rm LAO}$ and the sheet carrier density $\sigma_s$ with respect to LAO thickness are given in Fig.~\ref{fig:ER_vs_DFT}, as computed from first-principles and with the model based on the parameters $c$, $E_g^{\rm STO}$ and $\varepsilon_{r}^{\rm LAO}$. The overall agreement between the model and the DFT result is satisfying.

The main features of the Zener breakdown model can be summarized as follows:
i) below a LAO critical thickness $d_{c}^{\rm ZB}$, the interface is insulating, with the presence of a built-in field in LAO. The electrostatic potential drop across the LAO layer linearly evolves with the thickness of LAO;
ii) at a critical thickness $d_{c}^{\rm ZB}$, the surface ${\rm O}~2p$ valence states reach the  energy of the unoccupied ${\rm Ti}~t_{2g}$ states of the STO substrate, raised by the electrostatic potential: a Zener breakdown occurs, and the charges are transferred from the surface to the interface;
iii) above the critical thickness $d_{c}^{\rm ZB}$, the valence band edge of LAO is pinned to the bottom of the conduction band of STO by the charge transfer, and the band gap remains closed. This leads to the presence of a 2DES in STO, confined close to the interface, and the presence of holes at the -$\rm AlO_2$ surface of LAO.

The predictions of the model are in good agreement with first-principles results. Moreover, there is a large array of experimental results supporting the Zener breakdown hypothesis. Notably, the critical thickness has been consistently determined to be $4~{\rm u.c.}$ by several groups and methods, for films grown either from PLD~\cite{thiel2006} or MBE~\cite{segal2009} at high $p_{\rm O_2}$, with the contribution of $\rm V_O$ to the conductivity removed by annealing. 

Additionally, the sheet carrier density ($n_s \sim 4-9 \times 10^{13}~{\rm cm^{-2}}$ measured at low LAO thicknesses $< 8~\rm u.c.$) is in good agreement with the densities estimated from HAXPES~\cite{sing2009} and RIXS~\cite{berner2010} experiments at the same thicknesses. However, at higher LAO thicknesses, the measured carrier densities do not increase, in contrast with the predictions of the Zener breakdown model. There are other experimental evidences against this scenario. There has been mention of sizeable density of ${\rm Ti}~3d$-like states measured below the critical thickness (as early as 2~u.c. of LAO), with core-level spectroscopic measurements suggesting that the breakdown occurs almost immediately~\cite{sing2009, berner2010, takizawa2011}; however, as these charges remain trapped and do not contribute to interface conductivity, they may not originate from a polar catastrophe, and it is possible that they originate from oxygen vacancies buried in the STO substrate. In addition to the presence of sub-critical ${\rm Ti}~{3d}$ carriers, no mobile holes have been found at the LAO surface, and no hole states have been detected near the Fermi level~\cite{berner2013, plumb2017}. There have however been recent reports of the existence of a hole-sheet if the LAO is capped with STO, even at sub-critical LAO thicknesses~\cite{pentcheva2010,huijben2012, lee2018}. In References~\onlinecite{pentcheva2010} and \onlinecite{huijben2012} however, the holes are attributed to the ${\rm O}~2p$ states of the surface ${\rm TiO_2}$ layer, whereas in Reference~\onlinecite{lee2018}, the LAO interlayer thickness is larger than the threshold thickness value and the authors attribute the 2DHS to stem from the ${\rm O}~2p$ states of the $\rm AlO_2$ layer at the $p$-type interface with the capping layer. Finally, there has been reports of suppressed conductivity at any LAO thickness for samples grown at very high $p_{\rm O_2} (\sim$ $10^{-3} - 10^{-2}~{\rm mbar}$)~\cite{herranz2007, kalabukhov2011}. 

The Zener breakdown scenario, and its simulation from first-principles is a consequence of considering pristine systems, without any defects which may alter the electric field in the LAO overlayer, its dielectric properties, \emph{etc}. However, other compensation mechanisms may occur earlier than the Zener breakdown, which might explain some of the discrepancies between this simple, naive picture and the experiments.

\subsection{Electric-field driven surface redox mechanism}\label{section:surface_redox_model}

Until now, we have discussed how the electronic structure of STO/LAO heterostructure behave if no atomic reconstruction/defects occurs during growth, assuming a pristine heterostructure. STO is sensitive to doping, and donor impurities have been suspected to be at the origin of the 2DES at its interface. In fact, it is well known that La impurities~\cite{okuda2001} and oxygen vacancies~\cite{muta2005} act as $n$-type donors. The present Section focuses on the role of oxygen vacancies in such heterostructures.

Even if the STO substrate is insulating before the growth of the LAO epitaxy, it could be expected that the growth process induces oxygen vacancies. In the original paper~\cite{ohtomo2004}, it was already reported that the 2DES properties are affected by the growth conditions, in terms of mobility, sheet resistance and electron densities. Since then, there have been several studies focusing on the role of oxygen partial pressure ($p_{\rm O_2}$) during growth, as well as the effect of annealing on the 2DES properties~\cite{basletic2008,herranz2007,kalabukhov2007, siemons2007}. Three regimes have been identified: low $p_{\rm O_2}$ ($\sim$10$^{-6}~{\rm mbar}$), high $p_{\rm O_2}$ ($\sim$10$^{-4}~{\rm mbar}$) and very high $p_{\rm O_2}$ ($\sim$10$^{-2}~{\rm mbar}$). For samples grown at low $p_{\rm O_2}$ as in Reference~\onlinecite{ohtomo2004}, the sheet carrier densities are in the range $10^{14}-10^{17}~{\rm cm^{2}}$, with mobilies around $10^{4}~{\rm cm^{2}\:V^{-1}\:s^{-1}}$ and sheet resistance around $10^{-2}~\Omega$. For high $p_{\rm O_2}$, the carrier density is significantly reduced to $10^{13}-10^{14}~\rm cm^{-2}$, in better agreement with the polar catastrophe (at least, for low LAO thicknesses), and with the resistance increasing by a few orders of magnitude. Samples grown at low $p_{\rm O_2}$ have carrier densities around $10^{13}-10^{14}~\rm cm^{-2}$ if annealed after growth, suggesting that the carriers found in the unannealed low $p_{\rm O_2}$ samples originates from vacancies. Finally, samples grown or annealed at high $p_{\rm O_2}$ remain insulating.

These results question the validity of the Zener breakdown scenario, which cannot explain by itself the $p_{\rm O_2}$ dependence of the transport properties. 
{Moreover, they imply that the growth process triggers the conductivity, by inducing oxygen vacancies within the substrate of the STO. Growth at low $p_{\rm O_2}$ induces a 3D-like conductivity in samples~\cite{huijben2006}, which indicates that this might be the case. However, annealing processes suppress the 3D-like conductivity. Furthermore, the existence of a systematic critical thickness cannot be easily rationalized within such a scenario. It has been proposed that the origin of the carriers are not the vacancies in the STO substrate, but the vacancies that exists at the surface of the LAO films~\cite{li2011, bristowe2011, bristowe2014, yu2014}.}

Theoretical studies based on DFT~\cite{li2011,bristowe2011, yu2014} have considered the possibility of intrinsic doping from polarity induced oxygen vacancies at the LAO surface. {They highlight the possibility of an electric-field driven stabilization of vacancies at the LAO surface, which reconcile the existence of a threshold thickness as well as the sensitivity to conditions of growth. This process is different than as-grown creation of oxygen vacancies at the STO surface or in the LAO overlayer. In a sense, this scenario can still be considered as a polar catastrophe, even though the mechanisms behind the charge transfer differ from the Zener breakdown model.}
The first DFT studies~\cite{li2011,bristowe2011} exploring this hypothesis were performed in the GGA, which is known to underestimate band gaps and to predict spuriously the properties of defects~\cite{freysoldt2014}. The last study~\cite{yu2014} are based on a partial implementation of Hartree-Fock exchange (at fixed geometry, after relaxation using semi-local functionals), and goes beyond oxygen vacancies, considering other intrinsic defects. However, the study is limited to a single defects planar density, which can be expected to be far from the thermodynamical equilibrium, as argued in Refs.~\onlinecite{bristowe2011, bristowe2014}. {The present study reconsiders these theoretical developments: on one hand, in the calculations performed for this study, we fully relax the structures within the hybrid functional formalism, and we consider different densities of defects, at the limit of our computational capacity.}

\begin{figure*}
	\centering
		\includegraphics[width=1.\textwidth]{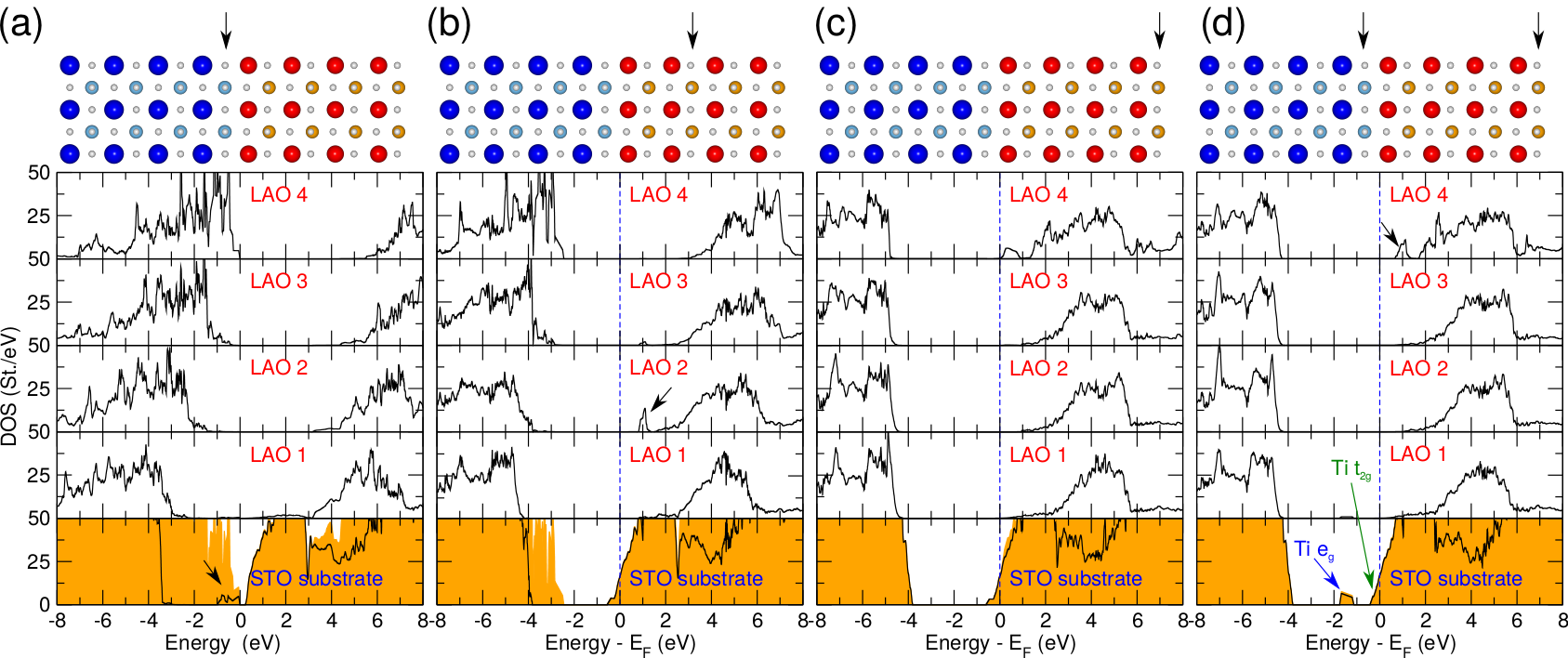}
        \caption{Layer-resolved DOS of ${\rm STO/LAO}_4$/vacuum heterostructures with $\rm V_O$ ($\eta = 1/4\square$) at different positions: (a) in the ${\rm TiO_2}$ layer at the interface; (b) in the middle of the LAO overlayer; (c) in the $\rm AlO_2$ layer at the surface; (d) in the ${\rm TiO_2}$ layer at the interface and in the $\rm AlO_2$ layer at the surface. The arrows indicate the positions of the $\rm V_O$ in the schematic representation of the heterostructures (top), and show the localized in-gap defect states in the densities of state.}
        \label{fig:slab4_4_2x2_VO_positions}
\end{figure*}

We first investigated the effect of oxygen vacancies considering a uniform distribution of oxygen vacancies at different positions in an otherwise pristine STO/LAO$_m$/vacuum heterostructure with a $n$-type interface, modelled through a symmetric slab as shown in Fig.~\ref{fig:HS}. For a $2 \times 2$ supercell containing one $\rm V_O$, the area density $\eta$ is equal to $1/4\square$. We first analyze the electronic structure of the defective system ${\rm STO/LAO}_4$/vacuum with oxygen vacancies ($\eta = 1/4\square$). Without any vacancies, this heterostructure is still predicted insulating within our formalism, just below the onset for Zener breakdown. It is therefore the ideal system to study different cases, based on the position of the vacancies. The layer-resolved DOS of such defective systems are given in Figs.~\ref{fig:slab4_4_2x2_VO_positions}.(a-d), for vacancies located (a) in the ${\rm TiO_2}$ layer at the interface; (b) in the middle of the LAO overlayer; (c) in the $\rm AlO_2$ layer at the surface; (d) in the ${\rm TiO_2}$ layer at the interface and in the $\rm AlO_2$ layer at the surface. 

It follows, based on these results:
(i) oxygen vacancies in the $\rm TiO_2$ layer at the interface are characterized by in-gap states slightly below the conduction band (up to $\sim$1~$\rm eV$), with a main ${\rm Ti}~3d_{3z^2 - r^2}$ character, which does not compensate the field in LAO as the donor electrons remain within the STO system;
(ii) if the oxygen vacancies are within the LAO layer, the defect in-gap states (shown with black arrows in Fig.~\ref{fig:slab4_4_2x2_VO_positions}) are always empty: they are always above the STO conduction band ($\sim$1~eV), hence the electrons will always be transferred to the interface {for any $\eta \leq 1/4\square$};
(iii) if the vacancies are at the LAO surface, the field is compensated over the whole LAO film. For vacancies buried within the LAO, the field will only be compensated between the interface the the LAO plane containing the vacancy;
(iv) as the defect states are empty if the vacancies are in the LAO layer, the $\rm V_O$ act as double donor. For $\eta = 1/4\square$, the carriers released by vacancies at the LAO surface completely compensate the field in LAO, the electrostatic potential is completely flat.

\begin{figure*}
	\centering
		\includegraphics[width=.75\textwidth]{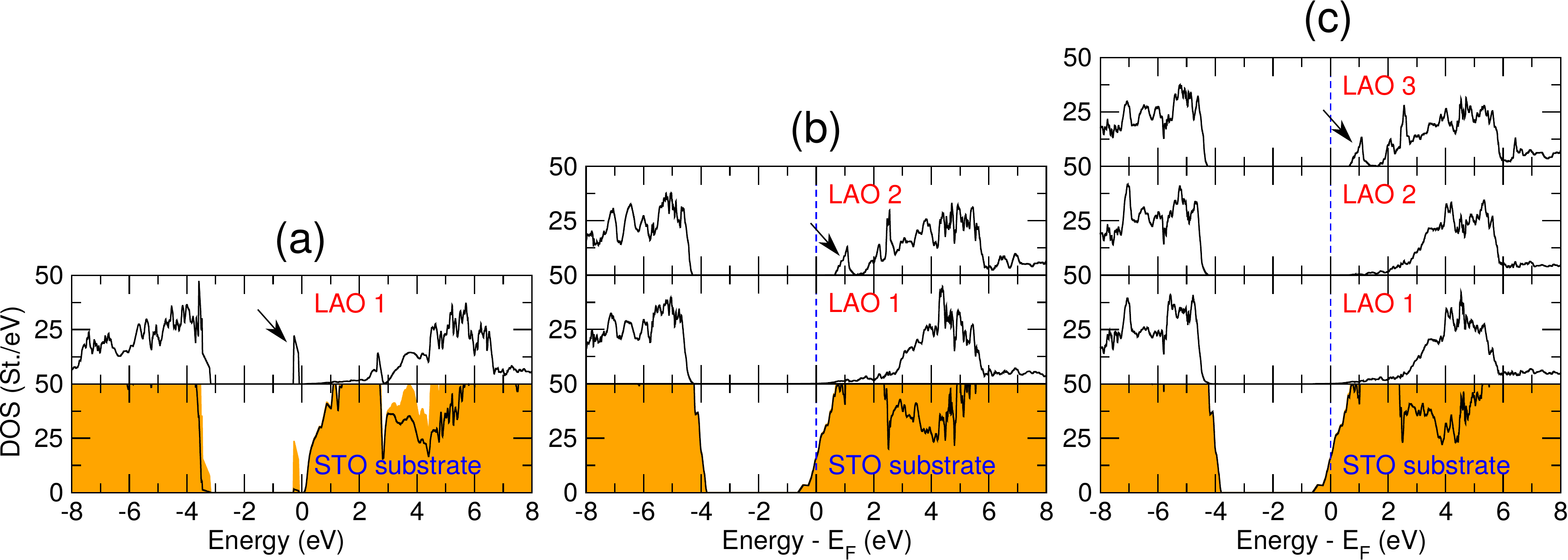}
        \caption{Layer-resolved DOS of ${\rm STO/LAO}_m$/vacuum heterostructures with $\rm V_O$ ($\eta = 1/4\square$) located in the $\rm AlO_2$ layer at the surface.}
        \label{fig:slab4_m_2x2_surface_VO}
\end{figure*}

The case where the vacancies are at the LAO surface is the most interesting case, since the carriers are transferred to the interface and they contribute the most to the screening of the built-in field. Figs.~\ref{fig:slab4_m_2x2_surface_VO}.(a-c) show the layer-resolved DOS for ${\rm STO/LAO}_m$ heterostructures with oxygen vacancies at the surface. For $m = 1$, the donor state is occupied, and below the conduction band of STO (hence, no charge transfer occur), for $m \geq 1$, the donor states are always above the conduction band, and charge transfer occurs. Hence, even in the absence of built-in field, it is always more favourable for the donor electrons to transfer to the interface rather than staying at the surface. More importantly, in contrast to the Zener breakdown scenario, oxygen vacancies at the LAO surface leave no mobile holes. 

\begin{figure*}
	\centering
		\includegraphics[width=1.\textwidth]{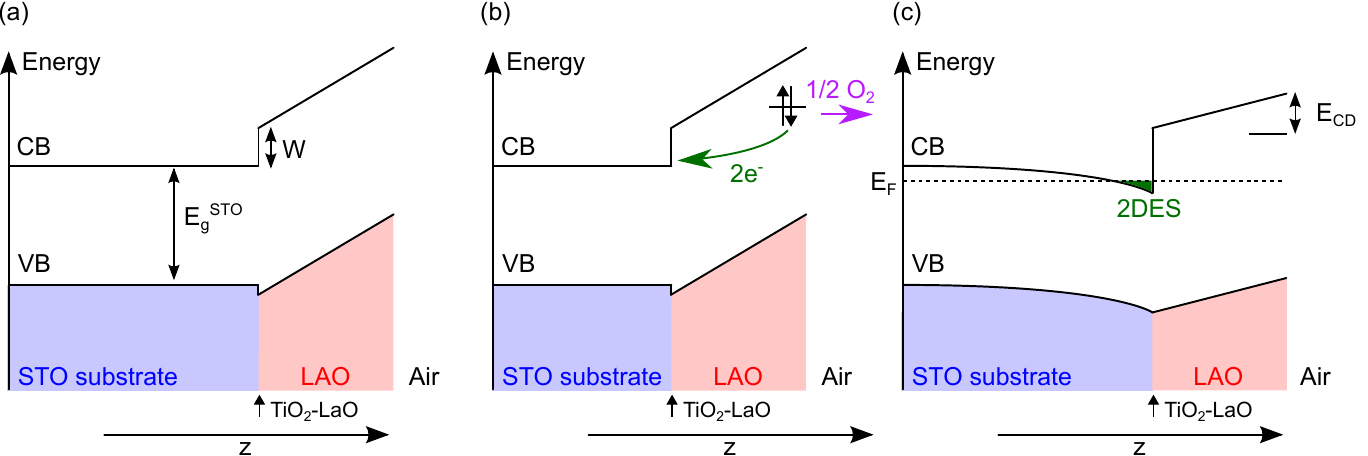}
        \caption[Schematic band diagram of the STO/LAO interface along the transverse direction]{Schematic band diagram of the STO/LAO interface along the transverse direction $z$: (a) the system below the critical thickness; (b) the creation of a double donor state through the formation of $\rm V_O$ at the surface. The donor electrons are transferred to the interface; (c) the charge transfer compensates the built-in field in LAO, depending on the charge density at the interface.}
        \label{fig:Surface_redox}
\end{figure*}

If we consider that each $\rm V_O$ at the LAO surface provide 2 electrons to the $n$-type interface, it is then interesting to know if the formation of these defects can be stabilized by the built-in electric field, and for which LAO thickness this stabilization occurs (it could occurs either before or after the threshold thickness for the Zener breakdown). The question has been addressed by Zhong \emph{et al.}~\cite{zhong2010} and Bristowe \emph{et al.}~\cite{bristowe2011, bristowe2014}, who build a generic model to understand the role of donor defects and how they may explain the experimental data where the Zener breakdown hypothesis fails. {The model of Zhong \emph{et al.}~\cite{zhong2010} is adapted for the superlattice geometry, whereas the one of Bristowe \emph{et al.}~\cite{bristowe2011, bristowe2014} account for heterostructures with a bare LAO surface, as in real samples.} We therefore aim to exploit this model, where the parameters are set by our predictions from first-principles, and to compare the results with the polar catastrophe expectations. The model is shown as a schematic representation in Fig.~\ref{fig:Surface_redox}, and is valid as long as the defect state is above the conduction band of STO (for $m > 1$).

The model consider the following: the formation energy of a single $\rm V_O$ at the surface of the film, $E_f(\eta)$, in the presence of an area density $\eta$ of surface $\rm V_O$, can be expressed as:
\begin{eqnarray}
E_f(\eta) &=& C + E_{\varepsilon}(\eta) + \alpha\eta\label{eqn:equation_model_first}
\end{eqnarray}
where $C$ is the energy cost of creating 1 $\rm V_O$ at the surface of LAO (in the absence of electric field), $E_{\varepsilon}(\eta)$ is the energy associated with the electric field in the polar layer, and the last term is a mean-field $\rm V_O$-$\rm V_O$ interaction (beyond electrostatics).

\begin{figure}
	\centering
		\includegraphics[width=1.\columnwidth]{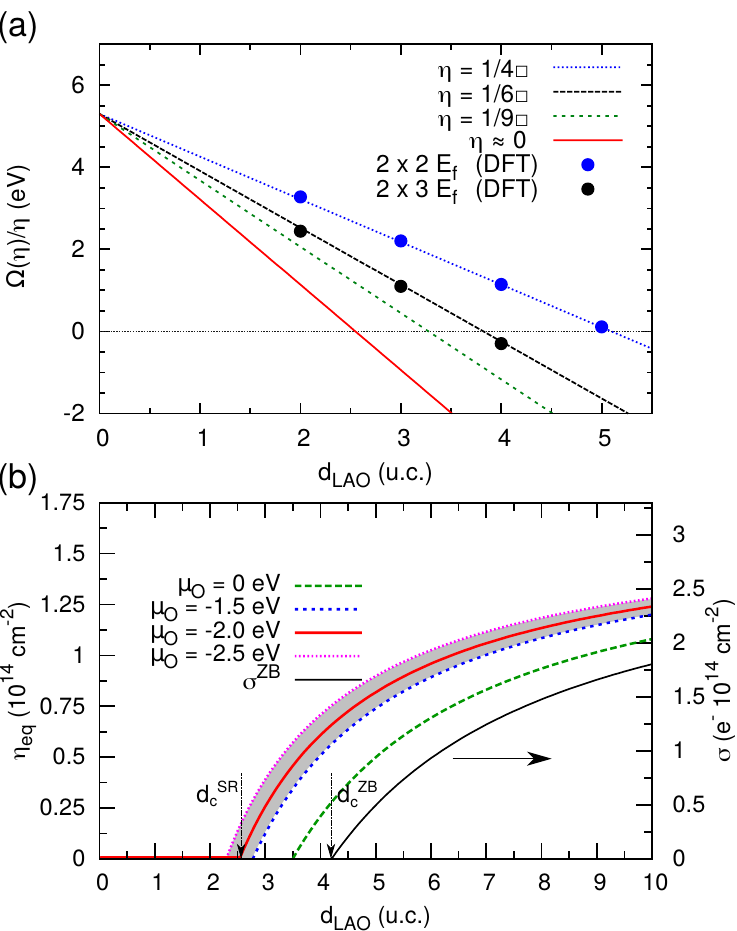}
        \caption{(a) Formation energies of $\rm V_O$ at the LAO surface versus LAO thickness $d^{\rm LAO}$ for different vacancy densities $\eta = 1/4\square$, $1/6\square$, $1/9\square$ and the limit toward $\eta = 0$, with parameters $\mu_{\rm O} = -2~\rm eV$ (see Appendix~\ref{AppendixA} for the details), $\varepsilon_{r}^{\rm LAO} = 22$, $C = 5.3~\rm eV$ and $\alpha = 0~\rm eV\:\angstrom^2$. The dots are the values calculated from DFT using a $2\times 2$ (blue) and $2\times 3$ (black) supercells; (b) equilibrium density of $\rm V_O$ with respect to LAO thicknesses for different values of the chemical potential. On the right axis, the corresponding carrier density at the interface if all the transferred charges contribute to transport ($\sigma = Z \eta$), and the predicted carrier density calculated within the Zener breakdown model. The arrows show the threshold thicknesses predicted by the surface redox ($d_{c}^{\rm SR}$) or the Zener breakdown ($d_{c}^{\rm ZB}$) models. The grey area highlight the variability with $\mu_O$.}  
        \label{fig:surface_redox_results}
\end{figure}

From Equation~\eqref{eqn:equation_model_first}, we consider the surface excess energy $\Omega$, accounting for the built-in electric field and the presence of an area density of $\eta$ oxygen vacancies at the surface:
\begin{eqnarray}
\Omega(\eta) &=& \int_{0}^{\eta}E_f(\eta')\:d\eta'\nonumber\\
\Longleftrightarrow\:\Omega(\eta) &=& C\eta + \Omega_{\varepsilon ,\rm  LAO}(\eta) + \frac{1}{2}\alpha\eta^2 \label{eqn:Omega_equation}
\end{eqnarray}
where the term $\Omega_{\varepsilon,\rm LAO}(\eta)$ accounts for the gain of electrostatic energy subsequent to the charge transfer. The $C$ term is a chemistry related term and depends on the energy cost of breaking bonds and on the overall chemical process. We also include within that term the electronic energy gain of the electron transfer from the defect state to the ${\rm Ti}~t_{2g}$ states of STO, which depends on the defect state binding energy $E_{CD}$ and the conduction band offset $W$:
\begin{eqnarray}
C &=& E_{f,\mu}^0 -  Z(W - E_{CD}) \label{eqn:C_term}
\end{eqnarray}
where $E_{f,\mu}^0$ is the formation energy of one single $\rm V_O$ at the $\rm AlO_2$ surface of LAO, $W$ and $E_CD$ are defined in Fig.~\ref{fig:Surface_redox}, and $Z$ is the number of carrier released by a single defect. The term $\Omega_{\varepsilon ,\rm  LAO}(\eta)$ takes the following analytical form:
\begin{eqnarray}
\Omega_{\varepsilon,\rm LAO}(\eta) &=& \frac{d^{\rm LAO}}{2\varepsilon^{\rm LAO}}\left[(\sigma_{c} - \eta Ze)^{2} - \sigma_{c}^{2}\right] \label{eqn:Omega_LAO}
\end{eqnarray}
This term is basically the energy gain of discharging a capacitor, and depends explicitly on the amount of transferred donor electrons $Z$ per $\rm V_O$. The electron charge is $e$. $\sigma_c$ corresponds to the charge density at the interface required to cancel the built-in field: $\sigma_c = 1/2~e/\square$. The equilibrium density of $\rm V_O$ $\eta_{eq}$ is the one minimizing $\Omega(\eta)$:
\begin{eqnarray}
\displaystyle \left[\frac{\partial \Omega}{\partial \eta}\right]_{\eta = \eta_{eq}} &=& 0 \nonumber\\
\displaystyle \Longleftrightarrow  \eta_{eq} &=& \displaystyle \frac{d^{\rm LAO}\:Z\:e\:\sigma_c - C\:\varepsilon^{\rm LAO}}{\displaystyle {d^{\rm LAO}} (Ze)^2 + \alpha\:\varepsilon^{\rm LAO}}\label{eqn:equilibrium_eta_for_capped_systems}
\end{eqnarray}
The stabilization of $\rm V_O$ ($\eta_{eq} > 0$) occurs at a critical thickness $d_{c}^{\rm SR}$:
\begin{eqnarray}
\displaystyle d_{c}^{\rm SR} &=& \displaystyle  \frac{C\:\varepsilon^{\rm LAO}}{Z\:e\:\sigma_{c}}\label{eqn:redox_critical_thickness}
\end{eqnarray}
Above this critical thickness $d_{c}^{\rm SR}$, and for large values of $d^{\rm LAO}$, $\eta_{eq}$ converges toward $\sigma_{c}/Ze$, toward a complete screening of the LAO electric field. In contrast to the Zener breakdown scenario, the critical thickness does not depend explicitly on the value of the band gap (Equation~\eqref{eqn:zener_breakdown_dc}), the band gap however can affect $C$. The model can be compared to first-principles results by comparing the DFT formation energies $E_{f}^{\rm DFT}$ and the model $\bar{E}_{f}$:
\begin{eqnarray}
\displaystyle \bar{E}_{f} &=& \displaystyle  \frac{1}{\eta}\Omega({\eta}) = \displaystyle  \frac{1}{\eta}\int_{0}^{\eta}{E_f}\:d\eta' \nonumber\\
&=& C + \frac{1}{\eta}\Omega_{\varepsilon ,\rm  LAO}(\eta) + \frac{1}{2}\alpha\eta \label{eqn:comparison_data}
\end{eqnarray}
which is the energy difference between the system with a given density $\eta$ of surface $\rm V_O$ and the pristine system ($\eta = 0$), per surface $\rm V_O$, and is basically the quantity calculated from Equation~\eqref{eqn:formation_energy_formula}. The parameters of the model are set in the following way: the chemical potential $\mu_{\rm O}$ is set to $-2~\rm eV$ (see Appendix~\ref{AppendixA} for details), similar to the value used in Refs.~\onlinecite{bristowe2011, yu2014}; for the $\rm V_O$-$\rm V_O$ mean-field interaction, we do not find a significant difference in formation energy between oxygen vacancies at the surface of $\rm STO/LAO_1$ heterostructures at different densities $\eta$, hence we set $\alpha = 0~\rm eV\:\angstrom^2$. Given that the donor state is always above the conduction band minimum (for $m > 1$), we have $Z = 2$. These parameters set, $C$ and $\varepsilon_{r}^{\rm LAO}$ are calculated through a fitting procedure of the model on DFT formation energies, yielding $C = 5.3~\rm eV$ (slightly larger than the value of $\sim 4.8~\rm eV$ in Ref.~\onlinecite{bristowe2011}) and $\varepsilon_{r}^{\rm LAO} = 22$ (slightly inferior to the value estimated from the Zener breakdown model).
The comparison between the model and the DFT formation energies estimated from Equation~\ref{eqn:formation_energy_formula_mu1} are given in Fig.~\ref{fig:surface_redox_results}.(a), showing that the model is in satisfying agreement with our calculations, despite the high density of $\rm V_O$ in the simulation cell, as the heterostructures are modelled through $2 \times 2$ and $2 \times 3$ supercells ($\eta^{\rm DFT} = 1/4\square$ and $1/6\square$).

In the limit of low density, we predict that $\bar{E}_f = \Omega(\eta)/\eta$ becomes negative as early as $d^{\rm LAO} = 2.5~\rm u.c.$, which corresponds to the onset for stabilization of surface $\rm V_O$ as shown in Fig.~\ref{fig:surface_redox_results}.(b). Accounting for the band gap error contribution to the $C$ term, $d^{\rm LAO} = 2.1~\rm u.c.$. The equilibrium density of surface $\rm V_O$, $\eta_{eq}$, with respect to LAO thickness is given in Fig.~\ref{fig:surface_redox_results}.(b): we also consider a range of $1~\rm eV$ across $\rm \mu_O$ to account for variations between the different growth conditions available in the literature, however this does not significantly alter the predicted threshold thickness. The effect of post-growth annealing is to shift $\mu_O$ toward 0~eV, hence we also consider this case in Fig.~\ref{fig:surface_redox_results}.(b): the threshold thickness is shifted to $3.5~\rm u.c.$. In all cases, the model predicts the stabilization of surface $\rm V_O$ below the critical thickness for Zener breakdown:
\begin{eqnarray}
d_{c}^{\rm SR} < d_{c}^{\rm ZB} \label{eqn:}
\end{eqnarray}
This means that the redox process is energetically more favourable than the creation of an electron-hole pair across the LAO film: this occurs when $C < Ze\:\Delta$.
Similarly to the Zener breakdown model, and for $\alpha \approx 0$, the surface redox model predicts a $1/d^{\rm LAO}$ thickness dependence for the carrier density at the interface above the threshold thickness, if all the charges released by surface $\rm V_O$ contribute to transport.

{These results also highlight an important point about computing defect formation energies in these heterostructures: the calculated values from DFT, using Eqn.~\eqref{eqn:formation_energy_formula}, shows a large dependence on defect density $\eta$ and LAO thickness (Fig.~\ref{fig:surface_redox_results}.(a)). This apparent dependence is not an artefact: it results directly from the second term in Equation~\ref{eqn:Omega_equation} (and possibly the third), which is implicitely accounted for in the DFT calculations due to the systematic charge transfer (Fig.~\ref{fig:slab4_m_2x2_surface_VO}). This has a significant implication: the defect densities simulated in previous studies~\cite{zhong2010,li2011, yu2014} (and this one as well), are actually far from equilibrium, as shown in Fig.~\ref{fig:surface_redox_results}, and one should remain cautious when analyzing the DFT results.}

\subsection{Electric-field driven surface protonation}

Surface protonation is another process from which the interface may be doped: hydrogen atoms adsorbed at the LAO surface are known to modulate the charge density at the STO/LAO interfaces, as argued from experimental~\cite{kim2014, scheiderer2015, brown2016, zhang2018} and theoretical work~\cite{adhikari2016,piyanzina2019}. To adapt the surface redox model to the specific case of hydrogen adsorbates {resulting from water splitting (Eqn~\eqref{eqn:hydroxylation_reaction})}, using the same Equations as in the previous section. Specifically, the $C$ term will be different ($C^{\rm H}$) {as it} depends on the chemical process; the $Z$ term is now equal to 1, as hydrogen adsorbates acts as single donors. Finally, the $\alpha$ term must also be reconsidered to account for the interaction between adsorbates. 

In this Section, we are mainly interested in surface protonation with ambient humidity as the source of the donor defects. We consider the water splitting process as given in Eqn.~\ref{eqn:hydroxylation_reaction}, where both gaseous dioxygen and water partial pressures ($p_{\rm O_2}$ and $p_{\rm H_2O}$, respectively) are parameters accounted through their respective chemical potentials.

DFT calculations performed on $2\times 2$ supercells including 1 or 2 H adsorbates have been used to perform a fit of the model to the adsorption energies as calculated from DFT using Eqn.~\eqref{eqn:formation_energy_H_adsorbate}, in the same fashion as done in the previous Section for oxygen vacancies. 

\begin{figure}
\centering
\includegraphics[width=1.\columnwidth]{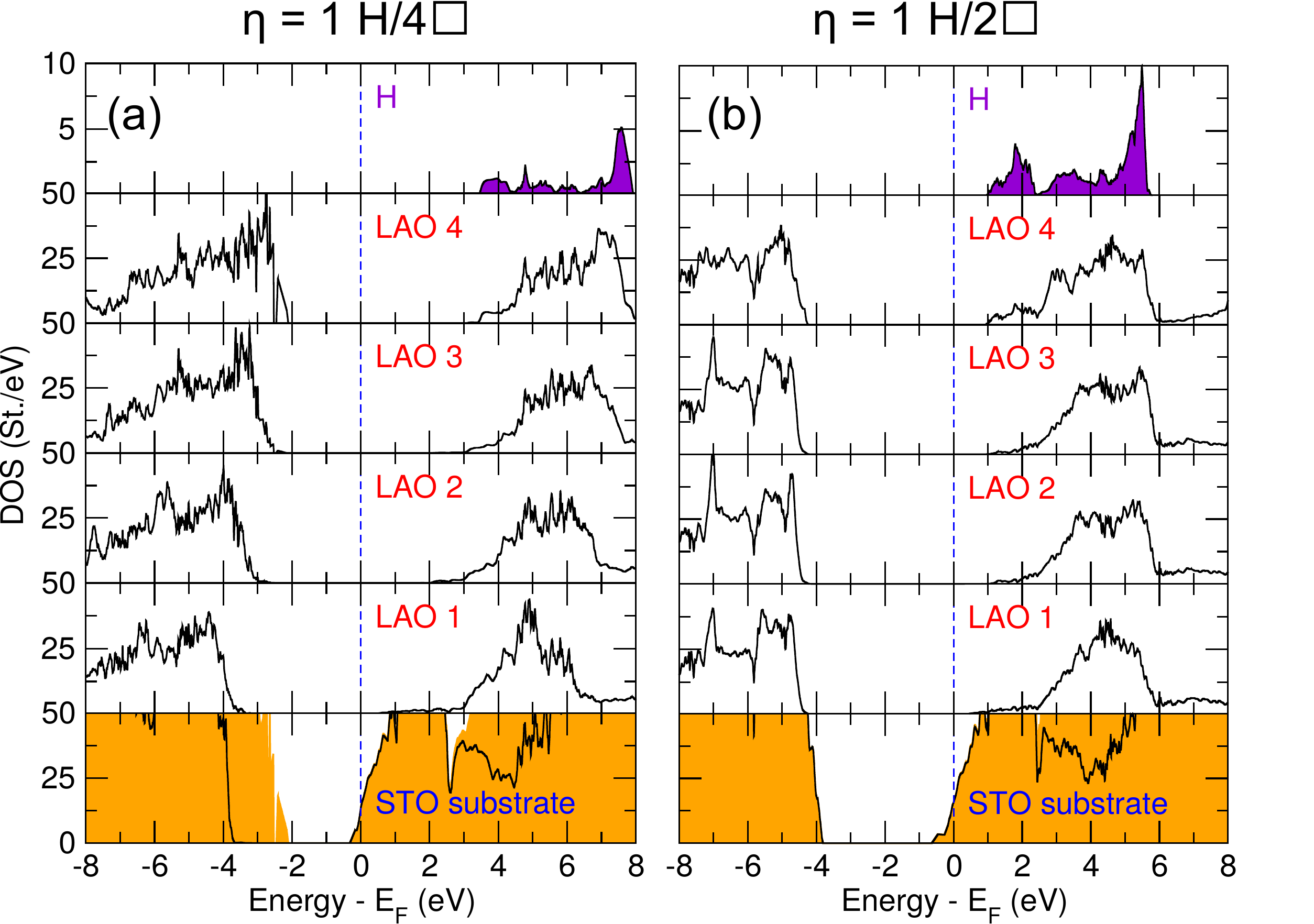}
\caption{{Layer-resolved DOS of ${\rm STO/LAO}_4$/vacuum heterostructures with H adsorbates at the surface: (a) $\eta = 1/4\square$; (b) $\eta = 1/2\square$.}}\label{fig:hydrogen_ads_dos}
\end{figure}

{The layer-resolved DOS for STO/LAO$_4$ with $\eta = 1/4\square$ and $\eta = 1/2\square$ are shown in Figs.~\ref{fig:hydrogen_ads_dos}. The DOS are similar overall to the ones with oxygen vacancies as surface defects (Figs.~\ref{fig:slab4_4_2x2_VO_positions} and~\ref{fig:slab4_m_2x2_surface_VO}), with the absence of a well defined donor state near the Fermi level. Electrons are directly transfered to the interface, and leaves no mobile hole at the surface. For the $\eta = 1/4\square$ case, there remains a non-zero electric field in the LAO overlayer.}

\begin{figure}
	\centering
		\includegraphics[width=0.93858260630188\columnwidth]{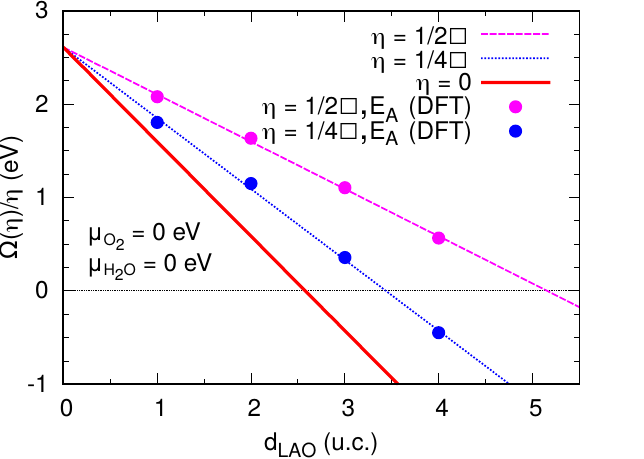}
        \caption{Adsorption energies of H adatoms at the LAO surface versus LAO thickness $d^{\rm LAO}$ for different adsorbate densities $\eta = 1/2\square$ and $1/4\square$, and the limit toward $\eta = 0$, with parameters $\mu_{\rm O_2} = \mu_{\rm H_2O} = 0~\rm eV$, $\varepsilon_{r}^{\rm LAO} = 22$, $C = 2.6~\rm eV$ and $\alpha = 0~\rm eV\:\angstrom^2$. The magenta (blue) dots are the values calculated from DFT using a $2\times 2$ supercell in the slab geometry with two identical surfaces and two (one) H per cell on each surface, using Eqn.~\ref{eqn:formation_energy_H_adsorbate}.}
        \label{fig:omega_fit_surface_protonation}
\end{figure}

The results are shown in Fig.~\ref{fig:omega_fit_surface_protonation}, for $\mu_{\rm O_2} = 0~\rm eV$ and $\mu_{\rm H_2O} = 0~\rm eV$. The fit results in $C^{\rm H}_{\mu = 0} = 2.6~\rm eV$, $\varepsilon_{r}^{\rm LAO} = 22$ and $\alpha = 0~\rm eV\:\angstrom^2$. Accounting for finite temperature and pressure, the $C^{\rm H}$ term is modified as follows:
\begin{eqnarray}
C^{\rm H}_{\mu} = C^{\rm H}_{\mu = 0} + \frac{1}{4}\mu_{\rm O_2} - \frac{1}{2}\mu_{\rm H_2O}\label{eqn:C_H_adsorbed}
\end{eqnarray}

\begin{figure}
	\centering
		\includegraphics[width=1.\columnwidth]{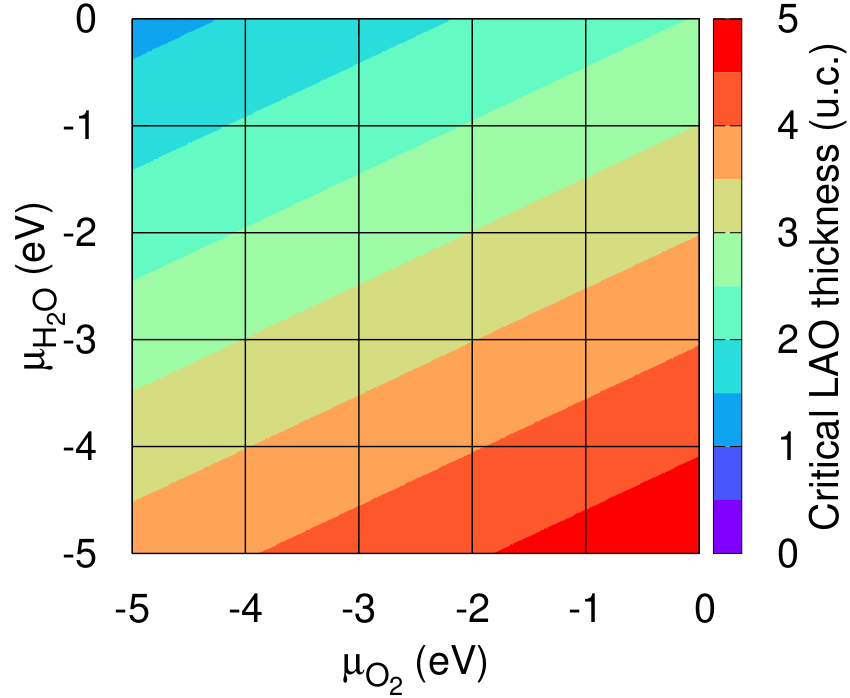}
        \caption{Critical LAO thickness for the surface protonation model, $d_{c}^{\rm SP}$, with respect to  $\mu_{\rm O_2}$ and $\mu_{\rm H_2O}$, as calculated from Eqn.~\eqref{eqn:critical_thickness_surface_protonation} using the parameters $C^{\rm H}_{\mu = 0} = 2.6~\rm eV$, $\varepsilon_{r}^{\rm LAO} = 22$ and $\alpha = 0~\rm eV\:\angstrom^2$.}
        \label{fig:critical_thickness_protonation}
\end{figure}

It is however difficult to determine the values of chemical potential for standard growth conditions of STO/LAO heterostructures, as $p_{\rm H_2O}$ is not usually provided in the literature. Nevertheless, using $\mu_{\rm O_2}$ and $\mu_{\rm H_2O}$ as free parameters, it remains possible to predict the critical thickness for exothermic surface protonation from:
\begin{eqnarray}
d_c^{\rm SP} &=& \frac{C^{\rm H}_{\mu}\varepsilon^{\rm LAO}}{Z\:e\:\sigma_c} \label{eqn:critical_thickness_surface_protonation}
\end{eqnarray}
where $Z = 1$. The critical thickness determined from Eqn.~\eqref{eqn:critical_thickness_surface_protonation} is shown in Fig.~\ref{fig:critical_thickness_protonation} with respect to $\mu_{\rm O_2}$ and $\mu_{\rm H_2O}$.
Using the upper bound of chemical potentials, the lower bound of LAO critical thickness is determined to be $2.6~\rm u.c.$. This is in agreement with the experiments of Scheiderer \emph{et al.}~\cite{scheiderer2015}, who managed to induce a metallic sheet conductance for an otherwise insulating STO/LAO$_3$ heterostructure through exposition to water vapor.
For typical growth conditions of STO/LAO heterostructures, the critical thickness is expected to be closer to 4~u.c. In a first approximation, given the similarities~\cite{chase1998} of standard entalphies  $H_0$ and entropy $S_0$ of dioxygen and water in gaseous form, then we can approximate $\mu_{\rm O_2} \approx \mu_{\rm H_2O}$ if the partial pressures of both gases are the same order of magnitude. Then $C^{\rm H}_{\mu} \approx C^{\rm H}_{\mu = 0} - \frac{1}{4} \mu_{\rm O_2}$. Hence, in the surface protonation picture, the effect of annealing is to shift $d_c^{\rm SP}$ to lower LAO thicknesses. This contrasts with the case of oxygen vacancies, {where it shifts} $d_c^{\rm SP}$ to larger LAO thicknesses. 

\subsection{Discussion}

In the thick layer limit ($d^{\rm LAO} \rightarrow \infty$), the Zener breakdown and the surface redox model predict the same charge density at the $n$-type interface. In the surface redox model, the potential drop is given by:
\begin{eqnarray}
V &=& (\sigma_c - \eta\:Z\:e)\frac{d^{\rm LAO}}{\varepsilon^{\rm LAO}}\label{eqn:surface_redox_potential_drop}
\end{eqnarray}
As we have $\alpha \approx 0$, and substituting $\eta_{eq}$ to $\eta$ above the threshold thickness, the potential drop across the LAO film is roughly equal to $C/Ze$, essentially independent of LAO thickness, as in the Zener breakdown scenario. The potential drop is pinned as the $\rm V_O$ are stabilized at the surface. 

The reduction in rumpling (cation-anion displacements) as measured by surface x-ray diffraction~\cite{pauli2011} and the sudden drop of the $c$-axis expansion above the threshold thickness~\cite{cancellieri2011} (reaching the elastic limit as early as $m = 6~\rm u.c.$), is achieved quicker than predicted by the Zener breakdown alone, suggesting an earlier onset for charge transfer, in agreement with the surface redox scenario. Additionally, no holes have ever been found at the LAO surface. The surface redox model is in better agreement with this observation, given that the donor states are $\sim$1~eV away from the Fermi level (pinned near the bottom of the STO conduction band) in absence of field, as in the fully compensated regime shown in Figs.~\ref{fig:slab4_m_2x2_surface_VO}(b-c). The $p_{\rm O_2}$ dependence of the sheet resistance also finds an explanation.

There remain open questions about the 2DES at the interface: if the origin of the carriers are $\rm V_O$, which arise for LAO thickness below $4~\rm u.c.$, why are the samples with smaller LAO thicknesses insulating ? It has been suggested that charge localization occurs through the Anderson mechanism~\cite{anderson1958}. In-gap states with a ${\rm Ti}~3d$ character have also been observed, at higher binding energies~\cite{berner2013,drera2011,koitzsch2011, ristic2012}, around $-1~\rm eV$ below the Fermi level. This description also fits the binding energy of the electron stuck in a $\rm V_O$ inside the STO substrate. Another possibility is that the growth process induces acceptor defects, such as cation intermixing, already observed in experiments~\cite{nakagawa2006, pauli2011, chambers2010, qiao2010, willmott2007, gunkel2010, kalabukhov2009, vonk2012}. It has also been suggested~\cite{bristowe2011} that such trapping states may be induced by the same donor $\rm V_O$ at the LAO surface: for low LAO thickness, surface $\rm V_O$ generate trapping potentials with a deep character, which become more shallow and numerous with increasing LAO thickness, releasing the carriers which may contribute to transport. In all cases, these hypotheses imply that the onset for conductivity is different than the onset for surface $\rm V_O$ stabilization. It is worth mentioning that our calculations involving $\rm V_O$ are still very close to pristine ${\rm STO/LAO}_m$ heterostructures, with a perfect interface, and no defects within the STO and LAO subsystems. Accounting for possible intrinsic defects change the results expected from a pristine interface, as studied in the present manuscript. The interfacial defects may be characterized as either deviations respecting the stoichiometry (inter-diffusion of atoms across the interface, such as $\rm Sr \leftrightarrow La$ or $\rm Ti \leftrightarrow Al$ ), and off-stoichiometric defects. The former is known to alter slightly quantitatively the expectations of the electric-field driven mechanisms, by inducing a potential shift at the interface~\cite{bristowe2014}, yet does not dope the interface~\cite{fontaine2014}, nor change the overall dielectric properties of the subsystems. The laters however may impact significantly the properties of the interface. A more detailed discussion can be found in Reference~\onlinecite{bristowe2014}. Finally, for heterostructures grown with a metallic capping layer characterized by a high work function $\phi_M$, x-ray photo-emission spectroscopy spectra~\cite{vaz2017} display signatures of metal oxidation, implying a chemical reaction between the LAO layer and the metallic capping layer. The degree of oxidation is also found to be correlated to the sheet carrier density. These results cannot be explained by the Zener breakdown alone, and are consistent with a redox process of the LAO surface: in this case, the oxygen atoms originating from the LAO remain trapped by the metallic layer, and the chemistry process and energetics will be different than that of the bare LAO surface process.
{Finally, from our analysis of the surface protonation process, it is likely that the shift of $p_{\rm O_2}$ does not simply change the critical thickness observed in experiments. Rather, a shift in $p_{\rm O_2}$ determines the dominant redox mechanism and bounds the values of the LAO critical thickness between 3 and 4~u.c.; this might explain the consistency of the values obtained in experiments, even if no Zener breakdown is occuring.}

\subsection{Tuning the polar discontinuity at oxide interfaces}

It has been shown that the threshold thickness can be tuned by replacing the LAO overlayer by an alloy made of STO and LAO, ${\rm Sr}_{1-x}{\rm La}_{x}{\rm Ti}_{1-x}{\rm Al}_{x}{\rm O}_3$, referred to as LASTO:$x$ in Reference~\onlinecite{reinleschmitt2012} where $x$ is the compositional \emph{ratio}. This observation can be rationalized within the electric field driven mechanisms. The rationale is that the formal polarization of the LASTO:$x$ overlayer $P^{0}_{{\rm LASTO:}x}$ can be changed continuously as:
\begin{eqnarray}
P_{0}^{{\rm LASTO:}x} &=& x\: P_{0}^{\rm LAO} \label{eqn:LASTOx_polarization}
\end{eqnarray}

if we assume a random alloying of both the $A$ (Sr and La) and the $B$ (Ti and Al) cations through the film so that $x$ ($1-x$) is the probability of occupation the La/Al (Sr/Ti) cations at a given $A$/$B$ site, in a virtual crystal approximation approach. Hence, for a composition $x = 0.5$, then the formal polarization of the alloy is half the formal polarization of the pristine LAO overlayer. 

\begin{figure}
	\centering
		\includegraphics[width=1.\columnwidth]{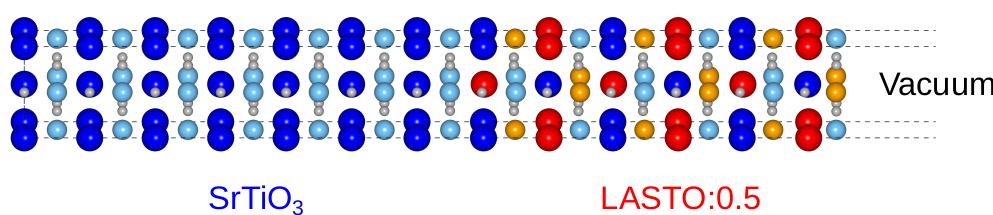}
        \caption[Geometry of the investigated STO/LASTO:0.5/vacuum system]{Geometry of the investigated system: the STO/LASTO:0.5/vacuum system is modelled in a symmetric slab geometry (only half the slab is shown for clarity), with a central off-stoichiometric STO layer and two LASTO:0.5 overlayers at each side of the slab, treated equivalently. The overlayer solid solution is modelled in a ``checkerboard'' configuration, alternating the cations along the transverse direction. Here the polar overlayer thickness is 6~u.c.}
        \label{fig:slab7_6_ALLOY_geometry}
\end{figure}

If the band gap of the alloy remains larger than the band gap of STO, the band alignment remains similar to that of the STO/LAO interface, and the dielectric properties of the polar layer close to that of LAO, the predicted threshold thickness for a Zener breakdown model becomes:
\begin{eqnarray}
d_{c}^{{\rm ZB, LASTO:}x} &=& \frac{1}{x}\: d_{c}^{\rm ZB,LAO}\label{eqn:LASTOx_critical_thickness}
\end{eqnarray}
Therefore, for a composition $x = 0.5$, the threshold thickness is expected to be twice the threshold thickness expected for the STO/LAO heterostructure. 
To investigate the specific case of LASTO:$0.5$, we performed calculations on heterostructures where the LAO overlayer is replaced by LASTO:0.5, with thicknesses $m = 1$ to 7~u.c.; the alloy is modelled as a homogeneous solid solution in a ``checkerboard'' configuration, as shown in Fig.~\ref{fig:slab7_6_ALLOY_geometry}. 

\begin{sidewaysfigure}
	\centering
		\includegraphics[width=1.\columnwidth]{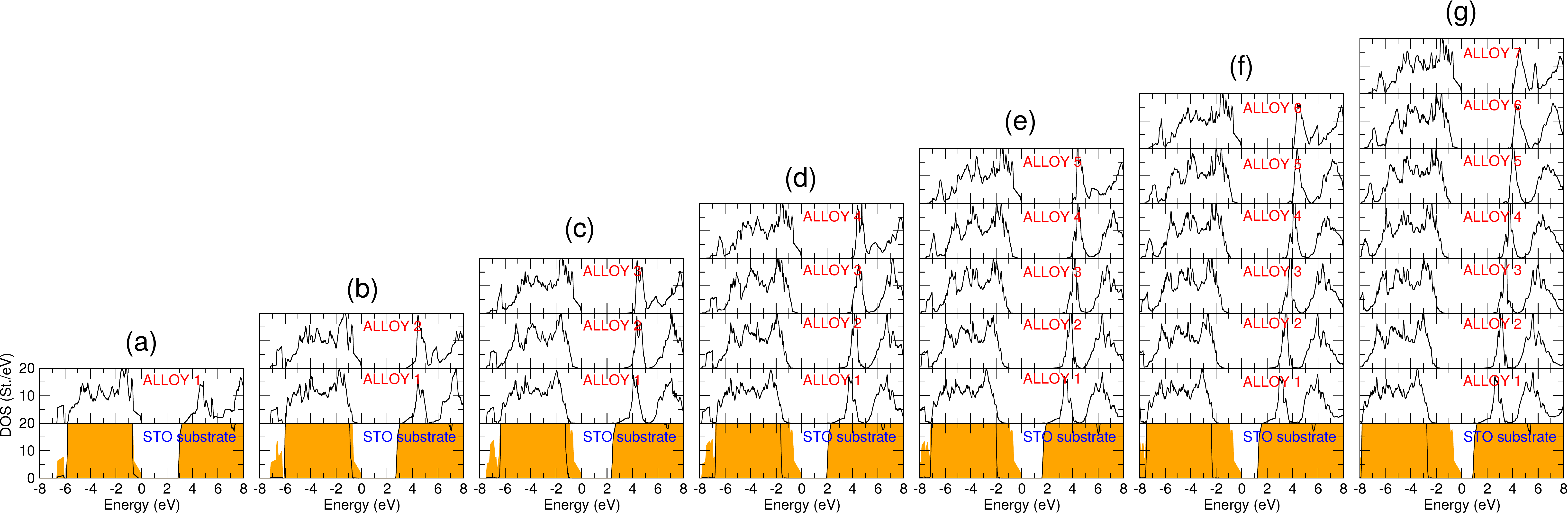}
        \caption[Layer-resolved DOS of ${\rm STO/(LASTO}$:$x)_m$/vacuum heterostructures]{Layer-resolved DOS of ${\rm STO/(LASTO}$:$x)_m$/vacuum heterostructures, for varying overlayer thicknesses $m$ (from $m = 0$ to $m = 7~\rm u.c.$). The orange area is the total DOS.}
        \label{fig:AlloySTO_slab7_m_DOSS_layer_resolved}
\end{sidewaysfigure}

The layer-resolved DOS for overlayer thicknesses up to 7~u.c. is given in Fig.~\ref{fig:AlloySTO_slab7_m_DOSS_layer_resolved}.(a-g): in all cases, the interface remains insulating, and we observe a linear decrease of the band gap, in agreement within the Zener breakdown scenario. 

\begin{figure}
	\centering
		\includegraphics[width=1.\columnwidth]{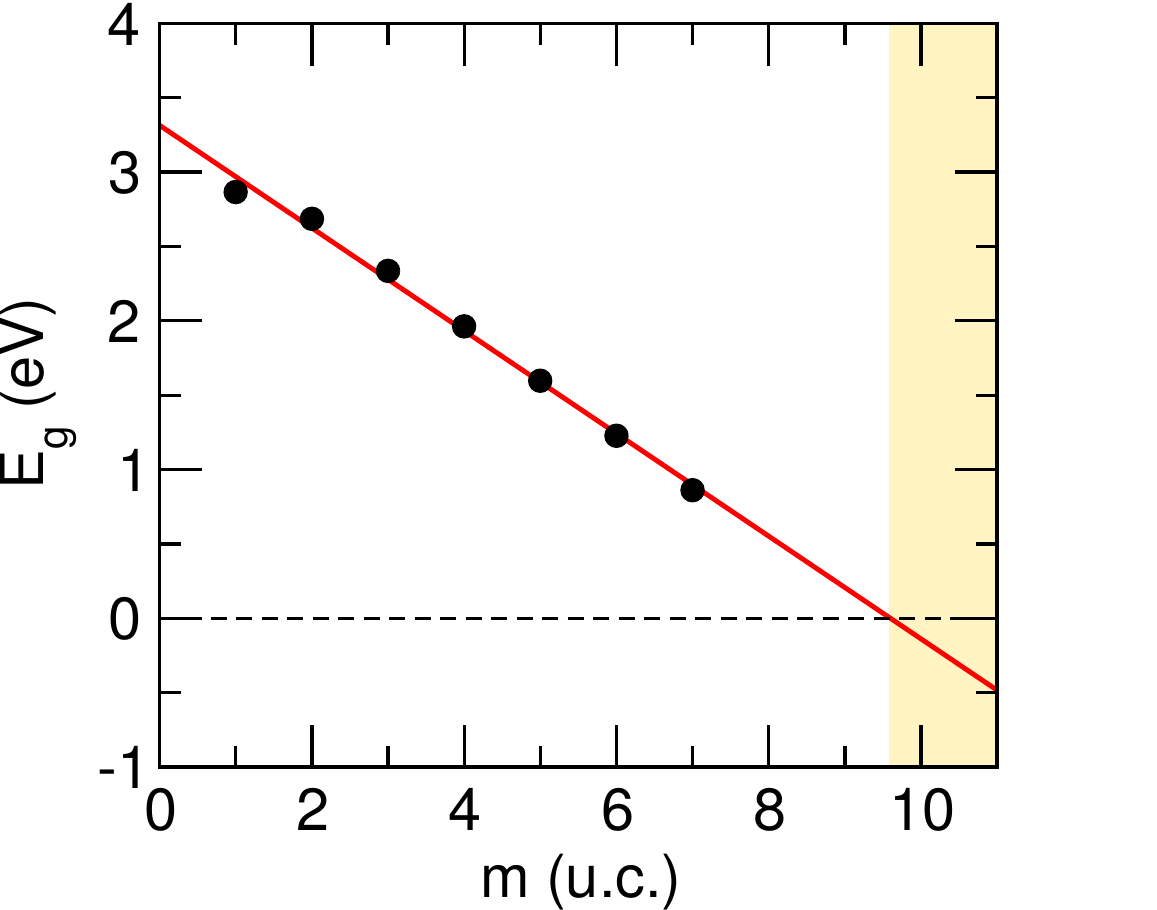}
        \caption[Electronic band gap for ${\rm STO/(LASTO}$:$0.5)_m$/vacuum heterostructures]{Electronic band gap for ${\rm STO/(LASTO}$:$0.5)_m$/vacuum heterostructures, for different thicknesses $m$ (u.c.), calculated as the difference between the bottom ${\rm Ti}~t_{2g}$ band and the top of the LAO ${\rm O}~2p$ band. The yellow area designate the thickness above which the interface is expected to be metallic as a result of a Zener breakdown.}
        \label{fig:AlloySTO_gap}
\end{figure}

Without performing an in-depth analysis as we did for the LAO case, by extrapolating the linear decrease of $E_g$ (Fig.~\ref{fig:AlloySTO_gap}), we estimate a critical thickness 
\begin{eqnarray}
d_{c}^{{\rm ZB, LASTO:}x} &=& 9.3~\rm u.c. \label{eqn:LASTOx_dc_DFT}
\end{eqnarray}
which is in agreement with the Zener breakdown model for $\varepsilon_{r}^{\rm LASTO:0.5} = 27$ (this value is confirmed by our hybrid functional \emph{ab initio} calculation of the bulk solid solution dielectric constant in the same atomic configuration). This corresponds to a built-in field of $\rm 0.11~V/\angstrom$ before the breakdown. 

The tunability of the critical thickness rationalized with the Zener breakdown scenario is in good agreement with the DFT calculations. Nevertheless, looking at the actual experimental~\cite{reinleschmitt2012,cancellieri2013} threshold thickness for $x = 0.5$, the MIT occurs between $5 - 6~\rm u.c.$, thus the Zener breakdown model overestimates the critical thickness (in our DFT calculations, the structure with $m$ = 7~u.c. is insulating, with a band gap of $0.86~\rm eV$), even when adjusting the potential drop to correct the overestimation of the STO band gap within the hybrid functional approach ($d_{c}^{\rm ZB} = 8.5~\rm u.c.$) as we did for the $x = 1.00$ case. It is therefore warranted to see to which extent the surface redox model predicts a threshold thickness in better agreement with the experimental value, {given that both models predict a composition dependence of the threshold thickness~\cite{bristowe2014}}.

\begin{figure}
	\centering
		\includegraphics[width=1.\columnwidth]{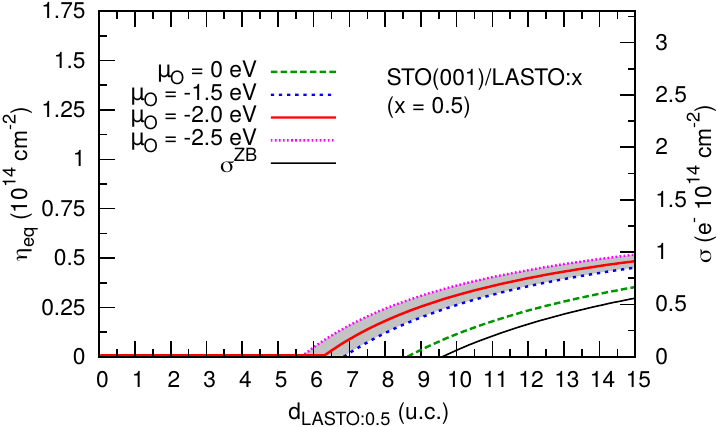}
        \caption[Equilibrium density of $\rm V_O$ at the surface of the LASTO:0.5 overlayer $\eta_{eq}$ calculated within the surface redox model]{Equilibrium density of $\rm V_O$ at the surface of the LASTO:0.5 overlayer $\eta_{eq}$ calculated within the surface redox model with parameters $C = 7.3~{\rm eV} + \mu_O$, $\varepsilon_{r}^{\rm LASTO:0.5} = 27$, $\sigma_{c} = 0.25~e/\square$, $\alpha = 0$. On the right axis, the corresponding electron density at the interface, with the prediction from the Zener breakdown model $\sigma^{ZB}$.}
        \label{fig:Alloy_surface_redox_results}
\end{figure}

For the case of $\rm V_O$ stabilized at the surface, let us first consider the $x = 0.50$ composition case. If we \emph{assume} that the chemical term $C$ remains close to the value calculated for the LAO surface, the equilibrium density of surface $\rm V_O$ $\eta_{eq}$ can be estimated for the alloy case by setting $\varepsilon_{r}^{\rm LASTO:0.5} = 27$ and $\sigma_{c} = xe/2\square = 0.25~e/\square$. For simplicity, we keep $\alpha = 0$ as for the $x = 1.00$ composition. The results are displayed in Fig.~\ref{fig:Alloy_surface_redox_results}: we predict a threshold thickness between 6 and 7~u.c. of LASTO:0.5, closer to the experimental result. In addition, we can see that the predicted surface density of $\rm V_O$ is lower than for the bare LAO case: only a transfer of $0.25~e^{-}/\square$ is required to completely screen the built-in field in the polar layer in this case. This value is however only reached in the infinitely thick limit. 

\begin{figure}
	\centering
		\includegraphics[width=1.\columnwidth]{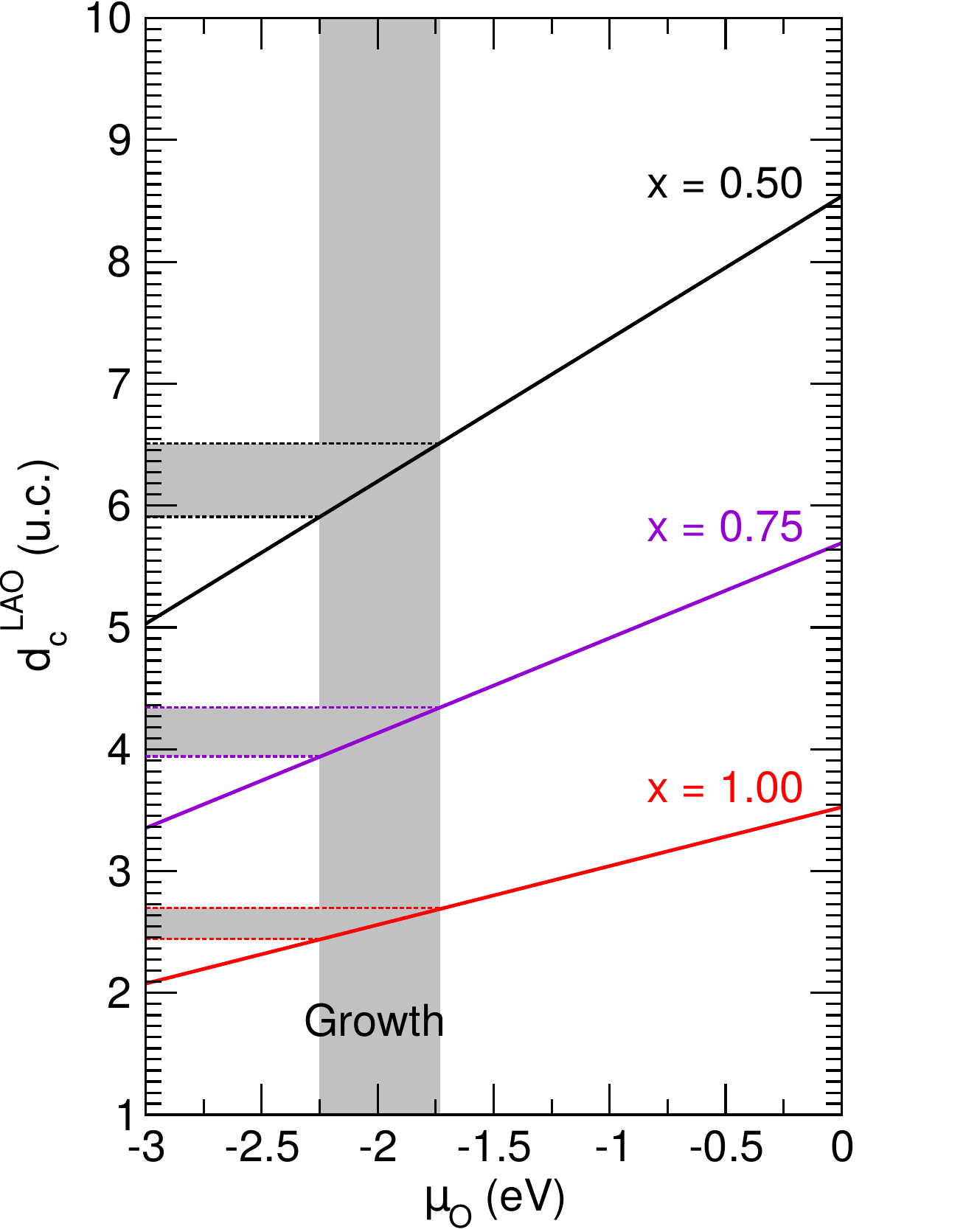}
        \caption[Polar layer threshold thickness of STO/LASTO:$x$ heterostructures as a function of chemical potential of oxygen $\mu_O$ as predicted by the surface redox model]{Polar layer threshold thickness of STO/LASTO:$x$ heterostructures as a function of oxygen chemical potential $\mu_O$ as predicted by the surface redox model $d_{c}^{\rm SR} = C\varepsilon^{{\rm LASTO:}x}/Ze\sigma_c$, estimated by fixing $C$ to its value calculated for the $x = 1.00$ composition (pure LAO overlayer). The grey band indicates the variation in $\mu_{\rm O}$ based on experimental setup for standard STO/LAO growth (see Fig.~\ref{fig:oxygen_chemical_potential}, Appendix~\ref{AppendixA}).}
        \label{fig:surface_redox_threshold_thickness}
\end{figure}

For $x = 0.75$, we predict a threshold thickness between 4 and 5~u.c., also in better agreement with the experimental value compared to the Zener breakdown scenario (6.1~u.c.). The threshold thicknesses for each composition are shown in Fig.~\ref{fig:surface_redox_threshold_thickness}, with respect to $\mu_{\rm  O}$. Of course, all of these predictions are based on the hypothesis that $C$ is the same in all cases, which remains to be proven; as the term is related to the chemistry of the surface, it is reasonable to expect some change. 

\section{Conclusions}

In this manuscript, we have re-examined the two most popular hypotheses for the origin of the carriers at the interface, namely the Zener breakdown and the polarity-induced surface redox mechanisms. The key physical parameters of these two models are notably well described with the B1-WC hybrid functional, which motivates the review of these models from first-principles with this novel approach. We show and discuss how the predictions of these models and \emph{ab initio} calculations compare with the experimental results. Our results indicate that oxygen vacancies and surface protonation at the LAO surface are typically stabilized at a lower LAO film thickness than the onset for Zener breakdown; this is related to the lower energy cost to form oxygen vacancies or Hydrogen adatoms at the LAO surface compared to the creation of an electron-hole pairs across the LAO film. Hence, for typical growth conditions of STO/LAO heterostructures, the Zener breakdown is unlikely to occur first. This justifies how some properties expected from the electron reconstruction, such as a metallic LAO surface, are not witnessed in experiments. Of course, the models and the first-principles calculations only involve pristine interfaces and surfaces. This may explain the failed predictions of the models with respect to some features of the 2DES witnessed in experiments, such as the sheet carrier density, typically overestimated within the models (for large LAO film thickness). The introduction of acceptor trapping states in these models is therefore worth investigating. 

Furthermore, we also discuss how the electric-field driven mechanisms may explain the experimental results obtained from STO/LASTO:$x$ heterostructures~\cite{reinleschmitt2012}, and highlight how the surface redox model may be more appropriate to explain the measured threshold thicknesses with respect to the composition of the polar layer.

\acknowledgements

We thank D. Fontaine, J.-M. Triscone, A. Filippetti, D. Li, S. Gariglio, M. Gabay and F. Ricci for fruitful discussions.
S.L. and Ph.G. were supported by the European Funds for Regional Developments (FEDER) and the Walloon Region in the framework of the operational program “Wallonie-2020.EU” (project Multifunctional thin films/LoCoTED). 
S.L. and Ph.G. were supported by the ARC project AIMED 15/19-09 785.
The present research benefited from computational resources made available on the Tier-1 supercomputer of the F\'ed\'eration Wallonie-Bruxelles, infrastructure funded by the Walloon Region under the grant agreement $\rm n^{\circ}~1117545$.


\appendix

\section{Chemical potential of oxygen}\label{AppendixA}

\begin{figure}
\centering
\includegraphics[width=1.\columnwidth]{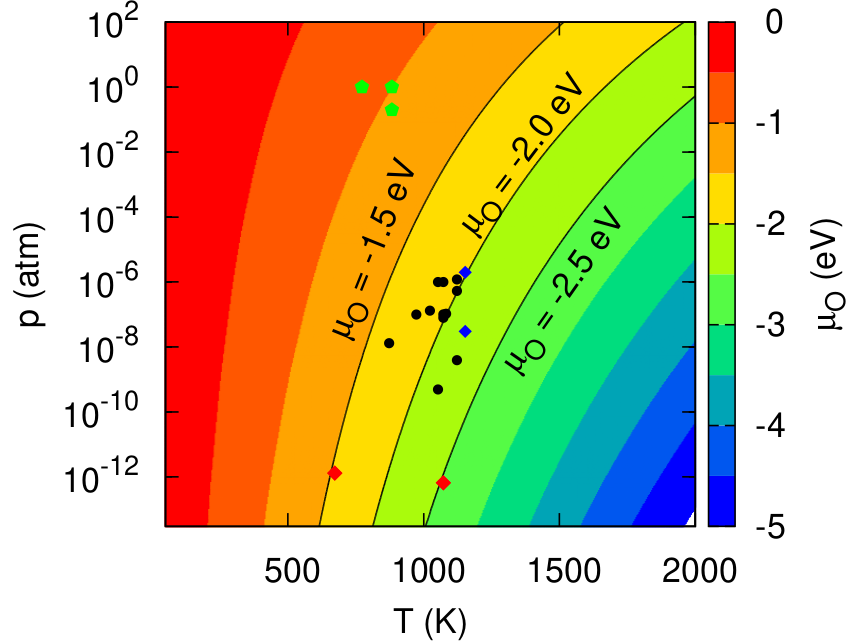}
\caption[Chemical potential of oxygen $\mu_O(p,T)$]{Chemical potential of oxygen $\mu_O(p,T)$, the black dots correspond to growth conditions extracted from the literature~\cite{drera2011, zaid2018,pfaff2018, takizawa2011, xue2017, li2018, cancellieri2013, salluzzo2013, pallecchi2010}. The two blue diamonds correspond to the growth conditions from Reference~\onlinecite{pentcheva2010}, for $p = 3\times 10^{-5}~\rm mbar$ and $p = 2 \times 10^{-3}~\rm mbar$ at $T = 1153~\rm K$. 
The green pentagons correspond to annealing conditions reported in References~\onlinecite{li2018, singh2018}, while the red diamonds correspond to the annealing conditions of bare STO films in ultra-high vacuum~\cite{dudy2016, cook2019} with a reported metallic surface.
}\label{fig:oxygen_chemical_potential}
\end{figure}

We calculate the chemical potential of oxygen at finite temperature and pressure from the thermodynamic model, inspired by the developments of References~\onlinecite{reuter2001,osorio-guillen2006}. Considering the environment as a gas reservoir of $N$ particles at pressure $p$ and temperature $T$, the chemical potential is given by the derivative of the Gibbs free energy:
\begin{eqnarray}
\displaystyle \mu = \left(\frac{\partial G}{\partial N}\right)_{T,p} = \frac{G}{N}
\end{eqnarray}
As $G$ is a potential function depending on $p$ and $T$, we can write, using the Maxwell relations:
\begin{eqnarray}
\displaystyle dG &=& \left(\frac{\partial G}{\partial T}\right) _{p}dT+ \left(\frac{\partial G}{\partial p}\right) _{T}dp\nonumber \\
 &=& -S\:dT + V\:dp
\end{eqnarray}
From the ideal gas equation of state ($pV = Nk_BT$), the partial derivative of $G(p,T)$ with respect to $p$ is:
\begin{eqnarray}
\displaystyle \left(\frac{\partial G}{\partial p}\right) _{T} = V = \frac{Nk_BT}{p}
\end{eqnarray}
In turn, a finite change of pressure from $p_0$ to $p$ results in:
\begin{eqnarray}
\displaystyle G(p,T) - G(p_0,T) &=& \int_{p_0}^{p}\left( \frac{\partial G}{\partial p}\right)_{T}dp\nonumber\\
 &=& Nk_BT\:ln\frac{p}{p_0}
\end{eqnarray}
Combining the first Equation and the last one, we have:
\begin{eqnarray}
\displaystyle \mu_{\rm O_2}(p,T) - \mu_{\rm O_2}(p_0,T) = k_B T\:ln\frac{p}{p_0}
\end{eqnarray}
Hence we have:
\begin{eqnarray}
\displaystyle \mu_{\rm O}(p,T) &=& \frac{1}{2} \mu_{\rm O_2}(p,T)\nonumber\\
 &=& \mu_{\rm O}(p_0,T) + \frac{1}{2} k_B T\:ln\frac{p}{p_0}\label{eqn:chemicalpotential}
\end{eqnarray}
From the knowledge of temperature-dependant $\mu_{\rm O}(p_0,T)$ at fixed pressure $p_0$ and Equation~\ref{eqn:chemicalpotential}, one can calculate the chemical potential $\mu_{\rm O}$ at given $p$ and $T$ using tabulated values for the $\rm O_2$ standard enthalpy $H_0$ and entropy $S_0$ (at $T_0 = 298~\rm K$ and $P_0 = 1~\rm atm$) through:
\begin{eqnarray}
\displaystyle \mu_{\rm O}(p_0,T) = &\displaystyle \frac{1}{2}& (\left[H_0 + \Delta H(T)\right] \nonumber\\
&-& T \left[S_0 + \Delta S(T)\right])
\end{eqnarray}
where $\displaystyle \Delta H(T) = C_p(T - T_0)$ and $\displaystyle \Delta S(T) = C_p\:ln\:\frac{T}{T_0}$. For typical growth conditions for the $\rm STO/LAO$ heterostructures, we use $p = 3.0\times 10^{-8}~\rm atm$ (oxygen partial pressure $p_{\rm O_2}$) and temperature $T = 1123~\rm K$~\cite{huijben2006}.
As for the standard values $H_0$, $S_0$ and heat capacity $C_p$, we use values from the NIST-JANAF thermochemical Tables~\cite{chase1998}:
\begin{eqnarray}
H_0 &=& 0~\rm kJ\:mol^{-1}\nonumber\\
S_0 &=& 205~\rm J\:mol^{-1} K^{-1}\nonumber\\
C_p &=& 29.39~\rm J\:mol^{-1} K^{-1}\nonumber
\end{eqnarray}
With these data, we find:
\begin{eqnarray}
H(T = 1123~\rm K) &=& 24.24~\rm kJ\:mol^{-1}\nonumber\\
S(T = 1123~\rm K) &=& 38.99~\rm J\:mol^{-1} K^{-1}\nonumber
\end{eqnarray}
This leads to $\mu_{\rm O}(p_0,\: T = 1123~{\rm K}) = -1.29~{\rm eV}$.
Finally, at pressure $p = 3.0\times 10^{-8}~\rm atm$:
\begin{eqnarray}
\displaystyle \mu_{\rm O}(p,T) &=& \mu_{\rm O}(p_0,T) + \frac{1}{2} k_B T\:ln\:\frac{p}{p_0}\nonumber\\
 &=& -2.13~{\rm eV}\nonumber
\end{eqnarray}
In the manuscript, we mainly use a rounded value of $\mu_{\rm O} = -2.0~\rm eV$, which is the same value used in Reference~\onlinecite{yu2014} and close to the value used by Bristowe \emph{et. al.}~\cite{bristowe2011} ($-1.9~\rm eV$). The chemical potential of oxygen with respect to $p$ and $T$ is given in Fig.~\ref{fig:oxygen_chemical_potential}, along with growth conditions ($p$,$T$) extracted from the literature, allowing us to determine the variation of $\mu_{\rm O}$ across the average value for the different experiments, determined to lie between $-2.2$~eV and $-1.7$~eV. The effect of post-growth annealing is to shift the chemical potential toward zero; depending on the annealing conditions, we expect the upper bound of chemical potential to be roughly $-1.0$~eV.

\newpage
\bibliographystyle{apsrev}
\setlength{\bibsep}{0.0pt}

\bibliography{Bibliography}

\end{document}